\documentclass[preprint,nofootinbib]{revtex4-1}
\usepackage{graphicx,epsfig,rotate}

\def\iso#1#2{\mbox{${}^{#2}{\rm #1}$}}
\def\he#1{\iso{He}{#1}}
\def\li#1{\iso{Li}{#1}}
\def\be#1{\iso{Be}{#1}}

\def\e10{\eta_{10}}

\def\neff{N_{\rm eff}}
\def\Ombh2{\Omega_{\rm b} h^2}
\def\omb{\omega_{\rm b}}

\def\avg#1{\langle #1 \rangle}

\def\ga{\mathrel{\raise.3ex\hbox{$>$\kern-.75em\lower1ex\hbox{$\sim$}}}}
\def\la{\mathrel{\raise.3ex\hbox{$<$\kern-.75em\lower1ex\hbox{$\sim$}}}}

\def\beq{\begin{equation}}
\def\eeq{\end{equation}}
\def\beqar{\begin{eqnarray}}
\def\eeqar{\end{eqnarray}}

\def\pfrac#1#2{\left( \frac{#1}{#2} \right)}

\def\bfrac{X}
\def\mfrac{{\cal X}}
\def\mfracX{{\cal X}}
\def\mfracY{{\cal Y}}

\begin{document}

\rightline{UMN--TH--3902/19}
\rightline{FTPI--MINN--19/25}
\rightline{December 2019}

\title{Big-Bang Nucleosynthesis After {\em Planck}}

\author{Brian D. Fields}
\affiliation{Departments of Astronomy and of Physics, University of Illinois,
Urbana, IL 61801}
\author{Keith A. Olive}
\affiliation{William I. Fine Theoretical Physics Institute,
School of Physics and Astronomy,
University of Minnesota, Minneapolis, MN 55455, USA}
\author{Tsung-Han Yeh}
\affiliation{Department of Physics, University of Illinois,~
Urbana, IL 61801}
\author{Charles Young}
\affiliation{Departments of Astronomy and of Computer Science, University of Illinois,~
Urbana, IL 61801}

\begin{abstract}
  We assess the status of big-bang nucleosynthesis (BBN) in light of
  the final {\em Planck} data release
  and other recent developments, and in anticipation of future
  measurements.
  {\em Planck} data from the recombination era
  fix the cosmic baryon density to 0.9\% precision,
  and now damping tail measurements
  determine the helium abundance and effective number of neutrinos with
  precision approaching that of astronomical and BBN determinations respectively.
  All three parameters are related by BBN.
  In addition, new high-redshift measurements 
  give D/H to better precision than theoretical predictions,
  and new Li/H data reconfirm the lithium problem.
  We present new $\be7(n,p)\li7$ rates using new neutron capture
  measurements; we have also examined the effect of proposed
  changes in the $d(p,\gamma)\he3$ rates.
  Using these results we perform a series of likelihood analyses.
  We assess BBN/CMB consistency, with attention to
  how our results depend on the choice of {\em Planck} data,
  as well as how the results
  depend on the choice of non-BBN, non-{\em Planck} data sets.
  Most importantly the lithium problem remains,
  and indeed is more acute given the very tight D/H observational
  constraints; new neutron capture data
  reveals systematics that somewhat increases uncertainty
  and thus slightly reduces but does not essentially change the problem.
  We confirm that $d(p,\gamma)\he3$ theoretical rates
  brings D/H out of agreement and slightly increases \li7;
  new experimental data are needed at BBN energies.
  Setting the lithium problem aside,
  we find the effective number of neutrino species at BBN is
  $N_\nu = 2.86 \pm 0.15$. Future CMB Stage-4 measurements
  promise substantial improvements in BBN parameters:
  helium abundance determinations will be competitive with
  the best astronomical determinations, and $\neff$
  will approach sensitivities capable of
  detecting the effects of Standard Model neutrino heating of the primordial plasma.
\end{abstract}

\pacs{}
\keywords{}

\maketitle

\section{Introduction}

It is remarkable that the two pillars of Big-Bang Cosmology, Big-Bang Nucleosynthesis (BBN) and the Cosmic Microwave Background (CMB)
have remained intertwined since the initial prediction of the CMB from the early
BBN studies of Alpher and Herman \cite{AH}. The reasoning behind the prediction of the CMB is relatively simple:
the Universe must have been hot
enough for nuclear reactions to proceed, which in the nascent Big Bang model meant within the first few minutes,
as determined via the Friedmann equation at BBN
\beq
\label{eq:Friedmann}
H^2 \approx \frac{8 \pi}{3} G_N \rho
\eeq
where $G_N$ is Newton's constant, and $\rho$ is the (radiation-dominated)
energy density
\begin{equation}
\rho \approx \rho_{\rm rad} = {\pi^2 \over 30} \left( 2 + {7 \over 2} + {7 \over 4}N_\nu \right) T^4 ,
\label{eq:rho}
\end{equation}
given in terms of the number of neutrino flavors, $N_\nu$.
Knowledge of the $p(n,\gamma)d$ cross section then allows for a determination of the baryon density, $n_B$ at that time.
Even a rough estimate of the baryon density today allowed Alpher and Herman to postulate the existence of the 
CMB with an estimated temperature of $T_0 =  5$ K. Going one step further, they recognized the potential uncertainty in
$n_B$, and argued that $T_0$ might be as low as 1 K. 

Seventy years after the initial prediction of the CMB, accurate parameter determinations from the CMB (in the $\Lambda$CDM
model) now are sensitive to both the primordial \he4 abundance and 
$N_\nu$. Accurate predictions of the light element abundances, in turn,
depend sensitively on the baryon density, now most accurately determined by CMB measurements. 
It should not be a surprise than that a full BBN study
requires a convolved likelihood analysis using CMB data along with the best available light element data.
Here, we update our previous analysis \cite{CFOY} (CFOY) using 2015 {\em Planck} data \cite{Planck2015}
with the latest release from {\em Planck} \cite{Planck2018}.

While not at the same level of precision as the CMB, BBN calculations and likelihood analyses
continue to make progress.  Since the standard model of BBN (SBBN) is well understood,
there continues to be room for improvement in cross section measurements,
the neutron mean life and of course light element abundance determinations. 
Since our analysis in CFOY, there have been some new cross section measurements,
notably in the $\be7(n,p)\li7$ rate that controls the $A=7$ abundance,
as well as controversy over the $d(p,\gamma)\he3$ discrepancy between
experiment and theory.
There have also been four new measurements of the neutron lifetime \cite{arz15,sere,pattie,ezhov},
and several new measurements of the deuterium abundance in high redshift quasar absorption systems 
\cite{riemer,bala,cookeN,riemer17,zava,CPS}.
There remains a discrepancy between the predicted \li7 abundance and that inferred
from observations of low metallicity halo dwarf stars,
further confirmed with a new observation of Li/H in an ultra-metal-poor
halo star \cite{aguado2019}. 

As in CFOY, we make heavy use of the public CMB likelihood distributions. 
Theses are combined with BBN likelihood functions which in turn rely on 
the observational determinations of the light element abundances. 
Our BBN likelihood functions convolve both theoretical and observational uncertainties. 
This enables us to construct both two-dimensional (as appropriate for SBBN with $N_\nu = 3$)
likelihood distributions for $\eta$ and the \he4 mass fraction\footnote{The \he4 mass fraction is defined and discussed in Appendix \ref{sect:convert}.}, $Y_p$, as well as three-dimensional 
distributions for $\eta$, $Y_p$, and $N_\nu$. When experimentally determined nuclear rates are 
used, we confirm the remarkable agreement between the CMB/SBBN predicted D/H abundance
with the now precise value from observation. However, we also comment on the consequences
of using a theory based rate for $d(p, \gamma)$\he3, normalized at low energies ($\sim 1$ keV) \cite{marc,coc15}.

Despite the quoted accuracy of the CMB parameter determinations, 
the CMB likelihood distributions clearly depend on the particular combination 
of CMB data used with or without other cosmological data. For example, {\em Planck} polarization (E-mode) data now places important constraints in addition to the usual temperature anisotropy data.  So the combination of temperature and E-modes auto correlations and cross correlations give the strongest CMB constraints for all parameters including those relevant to BBN.
Here, we consider several different {\em Planck} choices of data sets and compare results to test
BBN results to the sensitivity to {\em Planck} assumptions.

After a very brief description of SBBN, in \S \ref{sec:SBBN}, we describe the nuclear rates used in our BBN code,
and concentrate on rates which have been updated during the last three years. 
We also review the current status of neutron lifetime measurements.
Finally, in \S \ref{sec:SBBN}, we discuss the special case of $d(p,\gamma)$\he3 and prospects for improved
rates for this process. In \S \ref{sect:obs}, we update the status of \he4, D/H, and \li7/H determinations. We also describe in detail the set of {\em Planck} likelihood distributions we consider. 
Our BBN-only likelihood analysis is presented in in \S \ref{BBNlike}, and the convolution with the 
{\em Planck} likelihood functions for all data sets considered are given in \S \ref{sect:cmbpars}. Future prospects for improvement are discussed in \S \ref{S4}, 
and concluding remarks are made in \S \ref{sec:disc}.

\section{Standard Big-Bang Nucleosynthesis}
\label{sec:SBBN}

Until the age of precision cosmology, BBN depended primarily are three important input parameters \cite{ossty}:
the baryon number density $n_B$, the neutron mean-life, $\tau_n$,
and the number of neutrino flavors, $N_\nu$.
BBN thus directly probes the baryon mass density $\rho_B \approx m_B n_B$,
with $m_B \approx m_p$ mean mass per baryon \cite{gary}.
To compute the baryon density parameter $\Omega_B = \rho_B/\rho_{\rm crit}
= 8\pi G \rho_B/3 H_0^2$
engages the
infamous debate over the Hubble parameter $H_0  = 100 h \ \rm km/s/Mpc$;
this is avoided by quoting
$\omega_B \equiv \Omega_B h^2$.
One might also include the present temperature of the CMB which was known only to be between 2.7 and 3 K,
so that the baryon density can be expressed as the (present)
dimensionless baryon-to-photon ratio $\eta \equiv n_B/n_\gamma$.
In Appendix \ref{sect:convert}, we discuss the relation between $\eta$
and $\omega_B$.

Greatly improved measurements of 
the neutron mean-life, the accurate CMB determination of the baryon density first by WMAP \cite{wmap1}, and the LEP and SLC determinations \cite{lep}
of the number of neutrinos (in the standard model) from the invisible width of the Z boson, have effectively made BBN a zero-parameter theory \cite{cfo2}.
BBN results for the light element abundances of D, \he3, \he4, and \li7, are typically presented as a function of the 
baryon density (see for example the Schramm plot shown in Fig.~\ref{fig:schramm} below).
On the other hand, CMB measurements culminating in
{\em Planck} independently determine the cosmic baryon density
to better than 1 percent:
$\omega_B \equiv \Omega_B h^2 = 0.022298 \pm 0.000212$ \cite{Planck2018},
corresponding to $\eta \equiv n_B/n_\gamma = 6.104 \pm 0.058$.
This allows one to make firm predictions for the light element abundances.

For the most part, we will work in the context of SBBN, 
which adopts the most conventional microphysics and cosmology,
as follows.  We assume a
standard $\Lambda$CDM cosmology based on general relativity. We assume standard nuclear and particle
interactions, and use experimentally measured nuclear cross sections including the world average of the neutron mean-life.
SBBN will also refer to a particle content with three light
neutrino (and anti-neutrino) flavors that are left-handed and
fully in equilibrium.  

We will go beyond SBBN in the simplest and best-studied case
hereafter NBBN,
where we 
explore the effect of allowing $N_\nu$ to vary
in the cosmic radiation density (Eq.~\ref{eq:rho}).
Its effect is felt primarily through the expansion rate characterized by the 
Hubble parameter at the time of BBN (Eq.~\ref{eq:Friedmann})
Since the number of light neutrinos with weak charge is known to be 3,
$N_\nu > 3$, refers to the number of neutrino-like degrees of freedom of any new particle
type which may happen to be in thermal equilibrium at the time of BBN.
For example, a single light scalar would contribute 4/7 to $N_\nu$. 
For more on SBBN and NBBN, we refer to several works \cite{bbn,cfo1,bn,coc,cfo3,cuoco,cyburt,coc2,iocco,pisanti,coc3,cfo5,coc4,CFOY,coc18}.

\subsection{Updated Nuclear Rates}

Two nuclear reactions critical for BBN
have received particular attention since CFOY:
$\be7(n,p)\li7$ and $d(p,\gamma)\he3$.
Also, $\be7(n,\alpha)\he4$ is now measured for the first time, and there is new $\be7(d,p)\he4$ data.
Finally, the neutron mean life continues to 
be a source of controversy.

As we will see in more detail below, our approach to thermonuclear rates
has always been empirical, based on smooth polynomial
fits to experimental cross section data.
A complementary approach uses nuclear theory to prescribe fitting functions,
notably using $R$-matrix fits.  The latter has been explored
in detail by the Paris group and collaborators, e.g., refs.~\cite{coc2,coc15}.
A recent elegantly worked example of this approach
has been carried out for $\he3(d,p)\he4$
\cite{deSouza2018}.

Our choices for rates are those of~\cite{cyburt}, 
with modifications for those rates where new data exists in the BBN energy range.  In brief:  we use NACRE-II for the rates listed in Table \ref{tab:nacre2}.  We use \cite{ando} for 
$p(n,\gamma )d$.  And below we will discuss the remaining major rate, $\be7(n,p)\li7$, as well as the neutron lifetime, and other new rates for subdominant reactions.

\begin{table}[ht]
\caption{Reactions of relevance for BBN from the NACRE-II compilation \cite{nacreII}.
\label{tab:nacre2}
}
\vskip .1in
\begin{tabular}{|c|c|c|}
\hline
 $d(p,\gamma )\he3$ &
 $d(d,\gamma )\he4$ &
 $d(d,n)\he3$ \\
\hline
 $d(d,p)t$ &
 $t(d,n)\he4$ &
 $\he3(d,p)\he4$ \\
\hline
 $\li7(p,\alpha )\he4$ &
 $\li7(p,\gamma )\he4\he4$\ & \\
\hline
\end{tabular}
\end{table}

Also following~\cite{cyburt}, we model the uncertainty distribution as a lognormal distribution.  This is motivated physically by the idea that the experimental nuclear rates are controlled by several multiplicative factors whose uncertainties thus take this form \cite{coc15}.  In practice the errors are usually sufficiently small that the choice of a lognormal versus Gaussian distribution does not have a large impact on our result.

\subsection{The $\be7(n,p)\li7$ Reaction}
\label{sect:7benp}

The $\be7(n,p)\li7$ reaction was measured
at the n{\_}TOF experiment at CERN \cite{damone2018}.
The experiment used a neutron beam and a \be7 target, both of
which are radioactive.  This setup allows a direct measurement
of the reaction with forward kinematics,
i.e., via neutron-capture on \be7.
Most previous data is on the reverse reaction
$\li7(p,n)\be7$ that has a stable beam and target.
The two cross sections are related by time reversal invariance/detailed balance/reciprocity,
which gives
$\sigma(\be7+n)/\sigma(\li7+p) = E_{\rm cm}(\li7+p)/E_{\rm cm}(\be7+n)$.  Consequently, near the threshold energy for the reverse reaction, the transformation is very sensitive to the $Q$ value used to translate to the energy in the forward rate, and so at lower energies the forward rate data are particularly useful.

\begin{center}
  \begin{figure}[htb]
    \includegraphics[width=0.75\textwidth]{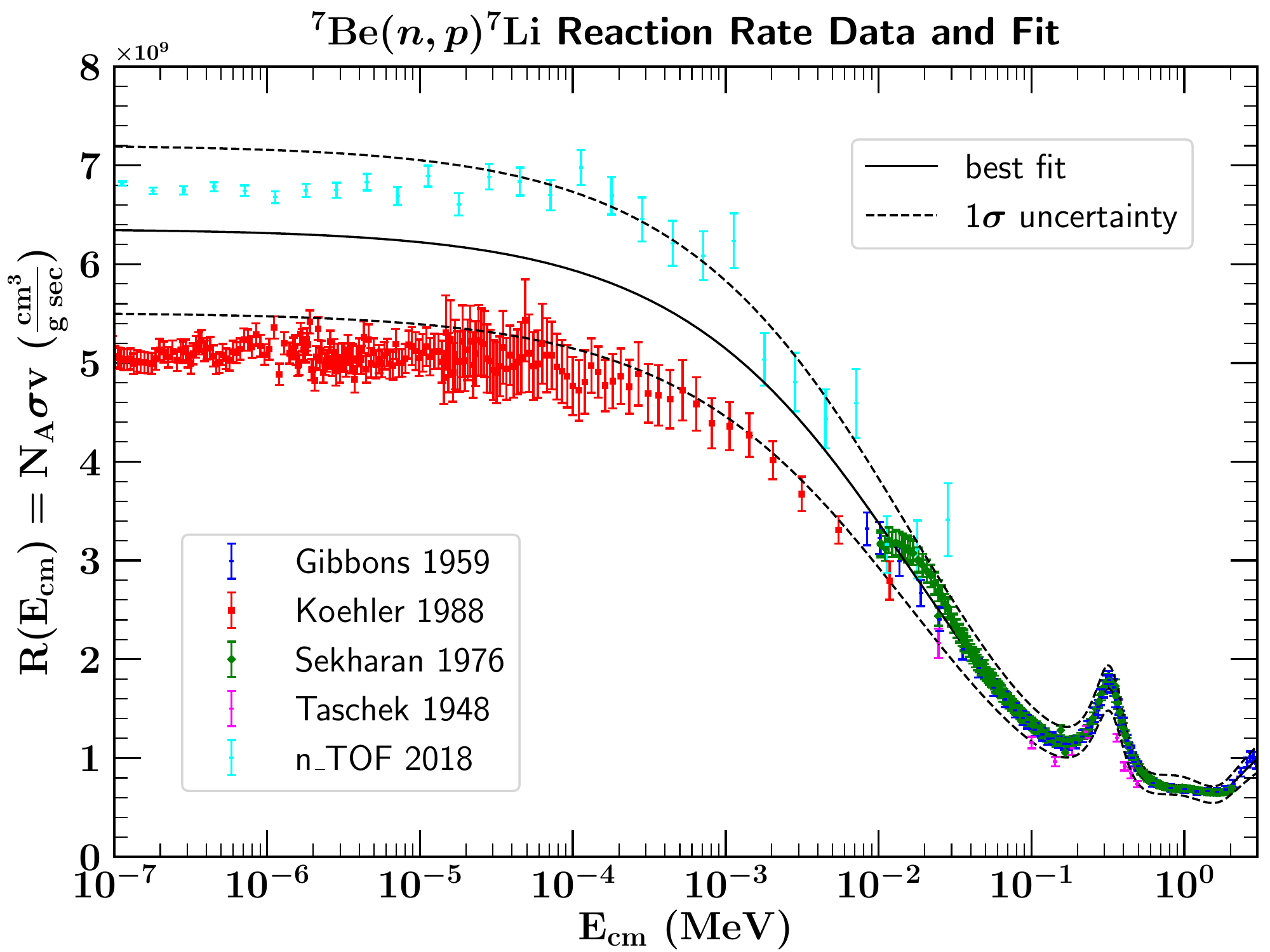}
    \caption{
      \label{fig:be7np}
      Data and fits for the $\be7(n,p)\li7$ reaction rate $R(E) = N_{\rm Avo} \avg{\sigma v}$ versus neutron kinetic energy $E$. Data from direct measurements are from {n\_TOF} \cite{damone2018} and Koehler \cite{koehler}.  Measurements from inverse kinematics are from Sekharan \cite{sekharan}, Gibbons \cite{gibbons}, and Taschek \cite{taschek}, and are transformed using detailed balance.  Note the resonances around 0.32 and 2.7 MeV; also note the discrepancies in the normalization at low energies.  Our polynomial fits are for different energy ranges and combinations of data sets, as described in the text.
    }
  \end{figure}
\end{center}

Figure \ref{fig:be7np} shows the data for $\be7(n,p)\li7$.
We see that the new n{\_}TOF \cite{damone2018} points are higher than those of
Sekharan et al.~\cite{sekharan} in and around the BBN energy range; this difference reaches substantially beyond the quoted errors.
The former are direct measurements of the forward reaction,
while the latter are from the reverse reaction, but we 
are not aware of a reason to exclude either.
Until a better understanding is forthcoming, this discrepancy
suggests that a systematic error may exist in one of the two
data sets.  Consequently, when we follow
the error propagation method of Cyburt \cite{cyburt},
the inclusion of n{\_}TOF will introduce a larger ``discrepancy error''
as well as a small increase in the mean cross section.  
At energies much lower than the BBN range, 
the n{\_}TOF data are also much higher than
the data of Koehler~\cite{koehler}, again suggestive
of systematic errors.

Our adopted reaction rate fits and thermally-averaged rates are
shown in Fig.~\ref{fig:be7np} 
and discussed further in Appendix \ref{sect:7benp-fits}.  
The fits include data in the energy range surrounding BBN energies,
$E_{\rm cm} \in (0.01,1.0) \ \rm MeV$.  In Appendix \ref{sect:7benp-fits}, we explore in detail the consequences of using or excluding different datasets and energy ranges for $\be7(n,p)\li7$.  

\subsection{Theory vs Experiment on $d(p,\gamma) \he3$}
\label{sect:dpg}

The $d(p,\gamma)\he3$ rate has received particular scrutiny lately because theory and experiment disagree on the behavior of this rate at BBN energies.  Our approach for this and all rates is to use smooth polynomial fits to the data.  By construction this ensures that the fits match the data as well as possible given the uncertainties.  On the other hand, this procedure makes no contact with nuclear theory.  

Nuclear theory predictions do exist for this reaction.  Such a 3-nucleon system is relatively straightforward to compute in {\em ab initio} quantum mechanical calculations based on nuclear interaction potentials normalized to nucleon scattering and other experimental data.   Nollett and collaborators first raised the issue \cite{Nollett2011,bbnt,cookeN} that in the BBN energy range, the theoretical prediction lies about 20\% above the data, which are mostly the Bailey and Ma  measurements \cite{bailey70,ma97}.  
Precision LUNA experiments are underway \cite{trezzi2018},
foreshadowed by a recent thick-target study \cite{tisma2019}.
Because the theoretical predictions do agree with the experimental data at energies above and below the BBN energy window, this raises a question of whether systematic errors in the experimental data could underestimate the cross section, as might be possible if the overall normalization is larger than the quoted 9\% error \cite{ma97}.

Other theoretical studies give similar results.  Another {\em ab initio} calculation based on
the nucleon-nucleon potential appears in \cite{marcucci2005}
and was updated in \cite{marc}, finding a higher cross section in the BBN range.
A theoretical fit for the cross section shape, based on an $R$-matrix analysis, 
was used by the Paris group \cite{coc15}.  This also gives a higher BBN cross section and rate.

To explore the impact of this discrepancy between theory and observation, we will show the impact of adopting a theory-based rate for this reaction. In addition to our fiducial results using empirical rates, we will adopt the theoretical rate of \cite{marc}.  This is similar that of the Paris group, but slightly higher in the BBN range.  And thus using this rate will illustrate the largest differences with our approach. Results based on the comparison of these rates will be given in Appendix \ref{sect:dpg-app}.

\subsection{Other Reactions:  Enhanced \be7 Destruction}

\label{sect:other-rates}

A competing channel for neutron capture on \be7
is $\be7(n,\alpha)\he4$, which leads to 
direct destruction of mass-7.
Data on this reaction has been scarce,
again due to the need for a radioactive source and target
for a direct measurement.
A recent flurry of experimental activity
by several groups has changed and clarified the picture
for this reaction.

The n\_TOF collaboration measured the rate at a wide
range of low
energies, from $10 \ \rm meV$ to $10 \ \rm keV$  \cite{Barbagallo2016}.
They determined the $s$-wave component,
$N_{\rm A} \avg{\sigma_s v} = 4.81 \times 10^{5} \rm \ cm^3 \ g^{-1} \ s^{-1}$.
Other measurements have included higher energies
than that spanning the BBN range.
Hou et al.~\cite{Hou2015} used detailed balance to
infer the cross section from the inverse $\he4(\alpha,n)\be7$ reaction.  
Kawabata et al.~\cite{Kawabata2017}
directly measured the forward reaction
using the neutron time-of-flight facility
at the Research Center for Nuclear Physics
in Osaka.  
Finally, Lamia et al.~\cite{Lamia2017} 
used the Trojan horse method to infer the reaction
properties indirectly, from $\li7+d$ deuteron breakup
and accounting for Coulomb effects.
This method is
well-suited to determine the {\em shape} of the cross-section but accrues more systematic errors when
inferring the absolute cross section.

Most importantly, all of the newly-determined $\be7(n,\alpha)\he4$ cross sections and thermal rates
were broadly consistent with each other (though significant
differences do exist). But they {\em all}
lie about an order of magnitude below the Wagoner \cite{wagoner} estimate that had been used  in the BBN code.  
Since even that higher
rate had not proven a dominant factor in BBN, we would
expect the revised rate to be even less significant--itself an important conclusion and useful
contribution.  
Ref.~\cite{Lamia2017} used the most recent data to
calculate a thermonuclear rate, and we adopt their
fit.
Below, in Appendix \ref{sect:other-rates-app},  we discuss the impact
of the new rate on primordial \li7.

A recent study of $\be7+d$ reactions found
a new resonance in the BBN Gamow window
for the $\be7(d,p \alpha)\he4$ channel \cite{Rijal2018}.
Resonances of this kind enhance mass-7 destruction, and
have been suggested as a possible solution
to the Li problem \cite{cp,chfo}.
Ref.~\cite{Rijal2018} performed a BBN study and found
a $\sim 10\%$ reduction in mass-7, more than
predicted in ref.~\cite{chfo}.
We will adopt their thermal rates and study the effect
of this reaction in our network. These results are also given in Appendix \ref{sect:other-rates-app}.

Finally, we note recent theoretical calculations on BBN rates.
A calculation of $\be7(\alpha,\gamma)\iso{C}{11}$
finds that it is unlikely this reaction proceeds through
a resonance strong enough to affect \li7 BBN production
\cite{hartos2018}.  
For the $\he3(\alpha,\gamma)\be7$ channel that dominates \li7 production, ref.~\cite{zhang2019} presents
a new, next-to-leading-order effective field theory calculation.
Their emphasis was on the $S(0)$ factor needed for solar neutrinos,
for which they found good agreement with recent inferences of this
quantity.

\subsection{The Neutron Mean Life}

As noted earlier, the neutron mean-life has seen dramatic improvements over the years,
but has become a source of controversy.
In 1980, the Review of Particle Properties \cite{rpp80} world average (using four measurements)
was $\tau_n = 917 \pm 14$ s. By 1990, the uncertainty improved to $\tau_n = 888.6 \pm \pm 3.5$ s \cite{rpp90},
based on seven measurements, though the drop in the uncertainty was largely due to the 
measurement by Mampe et al \cite{mampe} using ultra cold neutrons. 
The uncertainty hit a low in 2002 with the world average stabilizing at
$\tau_n = 885.7 \pm \pm 0.8$ s \cite{rpp02} for a period of ten years. The latter
average was based on six measurements all dating 1988 and later, and several with 
very small uncertainties including the measurement of Arzumanov et al. \cite{arz} of 
$\tau_n = 885.4 \pm 0.9 \pm 0.4$ s. All six measurements were within 1 $\sigma$
of the world average. However, the 2005 measurement of Serebrov  et al. \cite{sere05}
($\tau_n = 878.5 \pm 0.7 \pm 0.3$ s) using a gravitational trap was 6.5 standard deviations off from the world average
and was not included at that time. By 2012, two new measurements \cite{pilch,arz12} using ultra cold neutrons, 
also gave significantly lower values that the Review of Particle Physics average went down to
$\tau_n = 880.1 \pm 1.1$ s \cite{rpp12}, though some of the older measurements were somewhat discrepant with this values and
the uncertainty carried a scale factor of 1.8. The 2014 average was little changed \cite{rpp14} $\tau_n = 880.3 \pm 1.1$ s (scale factor 1.9)
and was the value adopted by CFOY.

In the 2018 compilation of the Review of Particle Physics \cite{rpp18}, the world average shifted slightly to 
$\tau_n = 880.2 \pm 1.0$ s due to the updated measurement in \cite{arz15} and includes a scale factor of 1.9
based on seven measurements.
As noted earlier, there have been several newer measurements  \cite{sere,pattie,ezhov}
and the world average based on all ten experiments gives 
$\tau_n = 879.7 \pm 0.8$ s with a scale factor of 1.9
due mainly to the high value found in \cite{yue}  ($887.7 \pm 1.2 \pm 1.9)$ s  and the low values found in 
\cite{sere,pattie} ($878.5 \pm 0.7 \pm 0.3$ s and  $877.7 \pm 0.7 +0.4/-0.2$ s).
The Review, however recommends dropping two of the older measurements (those in \cite{byrne,mampe})
as well as the in-beam measurement of \cite{yue}.
Thus the current world average becomes
\beq
\tau_n = 879.4 \pm 0.6  \ \rm s
\label{nlife}
\eeq
with a scale factor of 1.6 \footnote{The Review also quotes the average including \cite{yue} as
$\tau_n = 879.6 \pm 0.8$ s with a scale factor of 2.0.}.
Despite the variance between the in-beam and trap measurements~\footnote{See \cite{cms} for a review on this discrepancy.}, we adopt this latest world average in our
work below.

\section{Observations}
\label{sect:obs}

To test BBN and probe the early Universe, we require primordial abundance data to compare with our predictions.
With the exception of new CMB-measured \he4 (\S \ref{sect:CMB}), observed light element abundances in real astrophysical systems contain some non-BBN nucleosynthesis component,
so one must deduce the primordial abundances by removing or minimizing this contamination via observations of low metallicity and/or high-redshift systems.
For D, $\he4$, and $\li7$ it is possible to infer primordial levels at sufficient precision.  By contrast,
lack of low-metallicity observations of \he3 leaves us with no
reliable primordial abundances for this species \cite{he3}.
There is some hope, however, that future 30-m class telescopes
might detect \he3 emission lines in the same extragalactic HII regions
in which \he4 is now measured  \citep{he3cooke}, and to which we now turn.

\subsection{Helium-4}
\label{sect:he4}

As we will see below, the BBN prediction for the \he4 abundance is by far the most accurate of the 
light elements. Its relative uncertainty is at the level of 0.1\%. Therefore, a critical test of BBN
using \he4 would require a similar level of accuracy, which unfortunately is not possible at present.
\he4 abundance determinations are extracted from observations of extragalactic H II emission lines.
To obtain an abundance from a set of flux measurements requires a detailed model of the emission region,
atomic emissivities, and corrections for reddening and ionization fraction. Uncertainties in these 
lead to strong systematic uncertainties in the derived \he4 abundance \cite{osk,aos,aos3},
making better than 1\% determinations difficult \cite{osk,plp07}.
Despite the availability of order 100 observations of H II regions \cite{it,its07,isg13,itg},
only a few of these objects have data which are well described by the models \cite{aos,aos3}.
To compound matters, there are in general degeneracies in the parameters which 
make it difficult to determine abundances at high accuracy. 
However, recently calculated \he4 emissivities \cite{pfsd}  and the addition of a near infra-red line \cite{itg} have 
led to significant improvements in the \he4 abundance determinations \cite{aops,aos4}.

The near infra-red line helps break some of the degeneracies in the parameters used
to derive the \he4 abundance. 
In particular, its dependence on density and temperature differs from other observed He lines.
There are 16 objects satisfying $\chi^2 \la 6$ \cite{aos4} (roughly 2 $\sigma$ for two degrees of freedom). 
It was found \cite{aos4} that the inclusion of this line, reduced not only the uncertainty in the abundance determinations
of individual objects, but also the primordial abundance
\beq
Y_{p} = 0.2449 \pm 0.0040 + (78.9 \pm 43.3) {\rm O/H}
\label{ypr}
\eeq
inferred from a regression of the data to zero metallicity \citep{ptp74} with respect to oxygen (in units of $10^{-5}$). 
We use this value for $Y_p$ in our analysis below.

Recently there have complimentary analyses of \he4.
In \cite{ppl}, a select number of objects (5) were used and the slope of $\Delta Y/\Delta {\rm O} = 33 \pm 7$ was
assumed. They found $Y_p = 0.2446 \pm 0.0029$, where presumably the smaller uncertainty is related to the
smaller uncertainty assumed in $\Delta Y/\Delta \rm O$. Recent analysis \cite{ftdt} using SDSS-III data
and performing regressions against O/H, N/H and S/H found $Y_p = 0.245  \pm 0.007$.
It is encouraging that all three of these independent analyses agree well within the limits of their uncertainties.

\subsection{Deuterium}
\label{sect:deut}

It is well known that because of its steep dependence on the baryon density, the 
deuterium abundance is an excellent baryometer. Furthermore, 
since there are no known significant astrophysical sources for deuterium production \cite{els},
it is safe to assume that all observed deuterium is of primordial origin. While deuterium
can be destroyed through galactic chemical evolution, determinations of D/H at high redshift
help ensure that the observed abundance is close to primordial, particularly since most 
chemical evolution models indicate very little evolution of D/H at early times \cite{evol,opvs}.

There has been quite a lot of progress in D/H observations of quasar absorption systems. 
In CFOY, the adopted D/H value was based on 5 high quality observations yielding 
$ ({\rm D}{\rm H})_p = \left( 2.53 \pm 0.04 \right) \times 10^{-5}$ \cite{cooke}.
This result was derived from a 2012 observation of an absorber at $z = 3.05$  \cite{pc},
along with a subsequent new observation of an absorber at $z = 2.54$, and a reanalysis of
a selection of three other objects from the literature (chosen using a strict set of restrictions
to be able to argue for the desired accuracy)~\cite{cooke}. The relative uncertainties in D/H
for these systems are (2.1,2.8,4.1,6.0,6.0)\% (in order of increasing uncertainty).
Measurements with 2\% uncertainties are clearly remarkable. 

The last several years have seen several new D/H determinations.
These include a remeasurement of a high redshift system at $z = 3.36$ \cite{riemer}, though with a larger uncertainty of 11\%).
We include another system with relatively large uncertainty \cite{bala} of 15\%.
A very low metallicity system  (O/H = .0016 times the solar value) with a 2.8\% D/H determination was made available in 2016 \cite{cookeN}.
From 2017, we include three new observations; a very precise measurement (1.8\%) in a $z = 3.57$ absorber \cite{riemer17}, another 15\% measurement \cite{zava}
and a 3.4\% measurement \cite{CPS}. The weighted mean of these 11 measurements gives
\beq
\left( \frac{{\rm D}}{\rm H} \right)_p = \left( 2.55 \pm 0.03 \right) \times 10^{-5}
\label{d/h}
\eeq
Amazingly, this is very similar to the value adopted by CFOY with more than twice as many measurements. 
The variance in these measurements is also remarkably small, $0.18 \times 10^{-5}$ \footnote{The small, yet nonzero variance,
does lead to a slight scaling of the uncertainty in (\ref{d/h}). The unscaled uncertainty is 0.025 compared with the scaled uncertainty of 0.027
which we have rounded to 0.03.}.

\subsection{Lithium}

\label{sect:LiObs}

The lithium observational situation has changed little in recent years.
Milky Way halo stars are low metallicity sites for which lithium abundances
are available.  Stellar spectra show the neutral lithium
doublet at 6708\AA, but most of the Li atoms are singly ionized.
Thus, to find the star's present Li/H abundance requires most importantly
a model of the thermodynamic state of the atmosphere,
with results sensitive to the method of inferring the temperature,
and to assumption of local thermodynamic equilibrium
(LTE) \cite{hos,lind}.

Moreover, the possibility of stellar destruction of Li means
that the observations may not reflect the initial abundance.
Selection of candidates in the Spite plateau region of high
stellar temperatures \cite{spites}
selects stars with thin outer convective zones
where destruction should be minimal.  For the $> 100$ such stars found
at metallicities\footnote{Where $[{\rm Fe/H}] \equiv \log_{10}[\rm  (Fe/H)_{\rm obs}/(Fe/H)_\odot]$}
$-2.7 \la [{\rm Fe/H}] \la -1.5$, the Li/H abundance is uniform with small
scatter \cite{melendez2004}, consistent with expectations for a primordial abundance,
at a level \cite{sbordone2010,ryan2000}
\beq
\label{eq:Li_p}
\left( \frac{{\rm Li}}{{\rm H}} \right)_{p} = (1.6  \pm 0.3) \times 10^{-10} \, .
\eeq
As we will see in detail below, a primordial Li abundance as in (\ref{eq:Li_p})
is far below the CMB+BBN prediction; this is the well-known `lithium problem'
\cite{cfo5,fieldsliprob}.

However, at $[{\rm Fe/H}] \la -2.7$, the Li/H abundances develop a large scatter that seems to increase towards decreasing metallicity \cite{sbordone2010}.
The reasons for this `meltdown' are unclear, but apparently
some Spite plateau stars do destroy their lithium.
Whether this resolves the lithium problem hinges on the underlying
reason for the low-metallicity Li meltdown,
and whether the same physics--whatever that may
be--is also at play in stars not in the meltdown metallicity regime.
Also, even in the low-metallicity 'meltdown' region of high dispersion,
the Li/H abundances
do not significantly exceed the plateau value in Eq.~(\ref{eq:Li_p}).
Indeed, just recently lithium was found in an ultra-metal-poor star
which is carbon-enhanced but with iron 
unmeasurable and limited to $[{\rm Fe/H}] < -6.1$. 
In this remarkable object, lithium is found at 
${\rm Li/H} = (1.05 \pm 0.19)\times 10^{-10}$, below
but within $2\sigma$ of
the plateau \cite{aguado2019}.

\subsection{The Cosmic Microwave Background}

\label{sect:CMB}

The CMB provides an image of the universe at recombination.
The relevant causal region at that epoch
is the comoving sound horizon $r_{\rm s} = \int c_{\rm s} dt/a$,
which corresponds to angular scales $\theta_{\rm s} \sim 1^\circ$,
and multipoles $\ell \sim 200$.
Prior to recombination and on scales $< r_{\rm s}$,
gravity potential wells arising from density perturbations
drove motions in the cosmic baryon/photon plasma.
Radiation pressure countered these flows, leading to
(acoustic) oscillations that continued until photon decoupling.
Thus, small-scale CMB temperature and polarization anisotropies at angular
scales $\la \theta_{\rm s}$ record the acoustic oscillations at
the time of decoupling, with different modes in different phases.
The anisotropy power spectrum thus famously encodes a wealth of information
about the nature of the cosmos at recombination; for excellent
reviews see \cite{cmbrev}.

CMB anisotropy observations have achieved exquisite precision
in both temperature and polarization
anisotropies across
a wide range of angular scales.
This allows measurement of many cosmic parameters,
now including three that directly probe BBN:  (1) the baryon density
$\omb = \Omega_{\rm b} h^2$, (2) the helium abundance $Y_p$, and (3)
the number of neutrino species $N_\nu$.

Hou et al.~\cite{hou} elegantly lay out the influence of these
parameters in the CMB power spectrum, which we summarize briefly here.
The baryon density parameter $\omb$ has always been one of the
most robust and best-determined CMB outputs:
it controls the relative amplitudes of the alternating odd and even peaks,
which correspond
to modes undergoing maximal compression and rarefactions at the time of recombination.
More recently, CMB measurements have resolved small angular scales
where the fluctuations are suppressed by Silk damping -- photon
diffusion that smears the surface of last scattering \cite{silk}.  
The associated damping scale is the diffusion length
$r_d \sim \sqrt{Dt}$, where $t \sim 1/H$ and
the diffusion coefficient is set by
the Thompson scattering mean free path $D \sim \lambda_{\rm mfp} = c/n_e \sigma_{\rm T}$.  The number density of free electrons, $n_e$, excludes those
that have already recombined to He, and thus $n_e \sim (1-Y_p) \omb$.
Thus we have $r_d \sim 1/\sqrt{(1-Y_p)\omb H}$.

CMB observations
pin down the ratio of sound horizon to damping scales:
\beq
\label{eq:cmbratio}
\frac{r_{\rm s}}{r_{\rm d}} \propto \sqrt{\frac{(1-Y_p)\omb}{H}}
\propto \frac{(1-Y_p)\omb^{1/2}}{(1-N_\nu \rho_{1\nu}/\rho_\gamma)^{1/4}}
\eeq
where $\rho_{1\nu}$ is the energy density in a single light $\nu \bar\nu$ species.
With the ratio in Eq.~(\ref{eq:cmbratio}) fixed by CMB data,
we anticipate positive $(\omb,Y_p)$ 
and $(\omb,N_\nu)$ correlations, 
and a negative $(Y_p,N_\nu)$ correlation.  These
expectations are borne out in the results below.

In the {\em Planck} analysis, cosmic parameters are largely fixed by
CMB data, where constraints come from both temperature (T) and polarization
(E-mode) anisotropy measurements, in the form of TT,  EE, and TE power spectra.
One may choose to add constraints
such as 
large scale structure observations
and measures of the cosmic expansion rate and its history.
The {\em Planck} analysis presents a number of these possibilities, and
we will explore the effect on our analysis of adding different
non-CMB constraints.  Our baseline case will be the
\verb+Plik_TTTEEE_lowE_lensing+ analysis, which gives the
Markov chains representing the likelihood that combines
all of TT, TE, EE, low-$\ell$ reionization constraints, and CMB lensing
\footnote{\url{http://pla.esac.esa.int/pla/#cosmology}}.
This is also the fiducial case in the {\em Planck} cosmological
parameter study \cite{Planck2018}, and similar to the one used in our previous analysis \cite{CFOY}.

We will present {\em Planck} final CMB results for their two sets of analyses
in which results were found independently of BBN, i.e.,
without using BBN theory to relate $(\omb,Y_p,N_{\rm eff})$.
This allows us to test the CMB consistency with BBN.
The {\em Planck} chains
denoted \verb+base_yhe+ represent the likelihoods 
in $(\omb,Y_p)$ space while fixing $N_\nu = 3$. Converting the baryon density $\omb$ to 
the baryon-to-photon ratio, $\eta$, 
we denote this likelihood as 
${\mathcal L}_{\rm CMB}(\eta,Y_p)$ which is well-represented by
a 2D correlated Gaussian distribution.
The are however small perturbations
from a Gaussian and these are expanded by Hermite polynomials.  
For a more detailed description of the likelihood functions we use, see the Appendix in CFOY. 
We then relax this and allow $N_\nu$ to vary,
giving the \verb+base_nnu_yhe+ analyses. The three dimensional likelihood is denoted as ${\mathcal L}_{\rm NCMB}(\eta,Y_p,N_\nu)$.

The mean and standard deviation for our baseline case with $N_\nu = 3$ gives
\beqar
\omb^{\rm CMB} & = & 0.022298 \pm 0.000214 \\ 
\eta^{\rm CMB} & = & (6.104 \pm 0.058) \times 10^{-10} \label{2deta} \\
Y_p & = & 0.239 \pm 0.013  
\eeqar
Note that here and throughout, we determine the mean of $\eta$ and convert
between $\eta$ and $\omb$ using the relations
presented in Appendix \ref{sect:convert}.
The same Appendix discusses the small distinction between the \he4 {\em baryon} or nucleon fraction and {\em mass} fraction. {\em Planck} results present both mass and baryon fractions (their $Y_P$ and $Y_P^{\rm BBN}$ respectively); here and throughout we quote baryon fractions.
For comparison, the {\em Planck} 2015 results used in CFOY gave
$\omb = 0.022305 \pm 0.000225$
and $Y_p = 0.2500 \pm 0.0137$.

Allowing $N_\nu$ to vary, our baseline case gives mean and standard deviations of 
\beqar
\omb^{\rm CMB} & = & 0.022248 \pm 0.000220 \\ 
\eta^{\rm CMB} & = & (6.090 \pm 0.060) \times 10^{-10} \\
Y_{p,\rm CMB} & = & 0.246 \pm 0.019 \\ 
\neff & = & 2.859 \pm 0.314  
\eeqar
    {\em Planck} 2015 resulted in $\omb = 0.022212 \pm 0.00024$,
    $Y_p = 0.2612 \pm 0.0181$, and
    $\neff = 2.754 \pm 0.306$.
Below we will illustrate the degree to which these parameters are correlated.\footnote{Note that the values we quote here are using our Hermite polynomial fits to the {\em Planck} chains.  The resulting means and standard deviations can differ
from those directly from the chains
by less than one tenth of a percent in 2D fits. In our 3D fits, the differences in $\eta$ are also within 0.1\%, and within 1/2\% for
$Y_p$ and $N_\nu$.
The closeness of the values shows that our fits are excellent representations of the chains.}

While {\em Planck} is the culmination of a series of remarkably successful
space missions, future ground-based experiments
are envisioned as ``stage four'' of CMB science.
CMB-S4 promises improved
BBN parameters, particularly $Y_p$ and $N_\nu$.
As shown in ref.~\cite{CMB-S4,CMB-S4-RefDes}, the mission sky coverage particularly,
and also beam size, will determine the precision of these parameters;
forecasts span the ranges
$\sigma(Y_{\rm p}) \approx 0.0075 - 0.0040$,
and $\sigma(N_{\rm eff}) \approx 0.06 - 0.13$.
Indeed, accurate measurement of $\neff$ is a science driver for CMB-S4, with $\sigma(\neff)=0.030$ the target sensitivity.
If this can be realized,
then CMB-S4 should be able to resolve the $\neff - 3 = 0.045$ contribution from
neutrino heating in the Standard Model.
In \S \ref{S4} below we will consider the implications of these for BBN.

\section{The Likelihood Analysis and Monte-Carlo Predictions for the Light Element Abundances}
\label{BBNlike}

Over the last several decades, SBBN has evolved from a 2-parameter theory to essentially a parameter-free theory. The baryon density is now well defined with the statistically determined uncertainty. The neutron mean-life, despite current discrepancies, is quite well determined. As a result,
uncertainties in input BBN reactions
play a non-trivial role in determining
the uncertainties in the light element abundances and Monte-Carlo techniques \cite{kr,kk,hata,fo} have proven very useful and are now commonplace. 
Our procedure for constructing likelihood functions was discussed in detail in CFOY \cite{CFOY} and here we simply review the
necessary ingredients. 

We begin by computing the light element abundances and their uncertainties as a function of the 
baryon density. In the next section,
these likelihoods will be convolved with {\em Planck} chains and will allow more direct comparisons to observational data. 

We assume that each of the nuclear rates in our network have a lognormal 
distributed uncertainty and we randomly draw rates from those distributions for each value of 
$\eta$ (or $(\eta, N_\nu)$ when we allow the number of degrees of freedom to vary from the Standard Model value). For each input, the BBN code is run 10000 times establishing a mean
value and uncertainty for each of the light element abundances. We will denote the BBN likelihood functions
as ${\mathcal L}_{\rm BBN}(\eta;X)$
for  $X \in (Y_p,{\rm D/H},{\rm \he3/H},{\rm \li7/H})$, and 
${\mathcal L}_{\rm NBBN}(\eta,N_\nu;X)$ when $N_\nu$ is allowed to vary.

\begin{center}
\begin{figure}[!htb]
\includegraphics[width=0.7\textwidth]{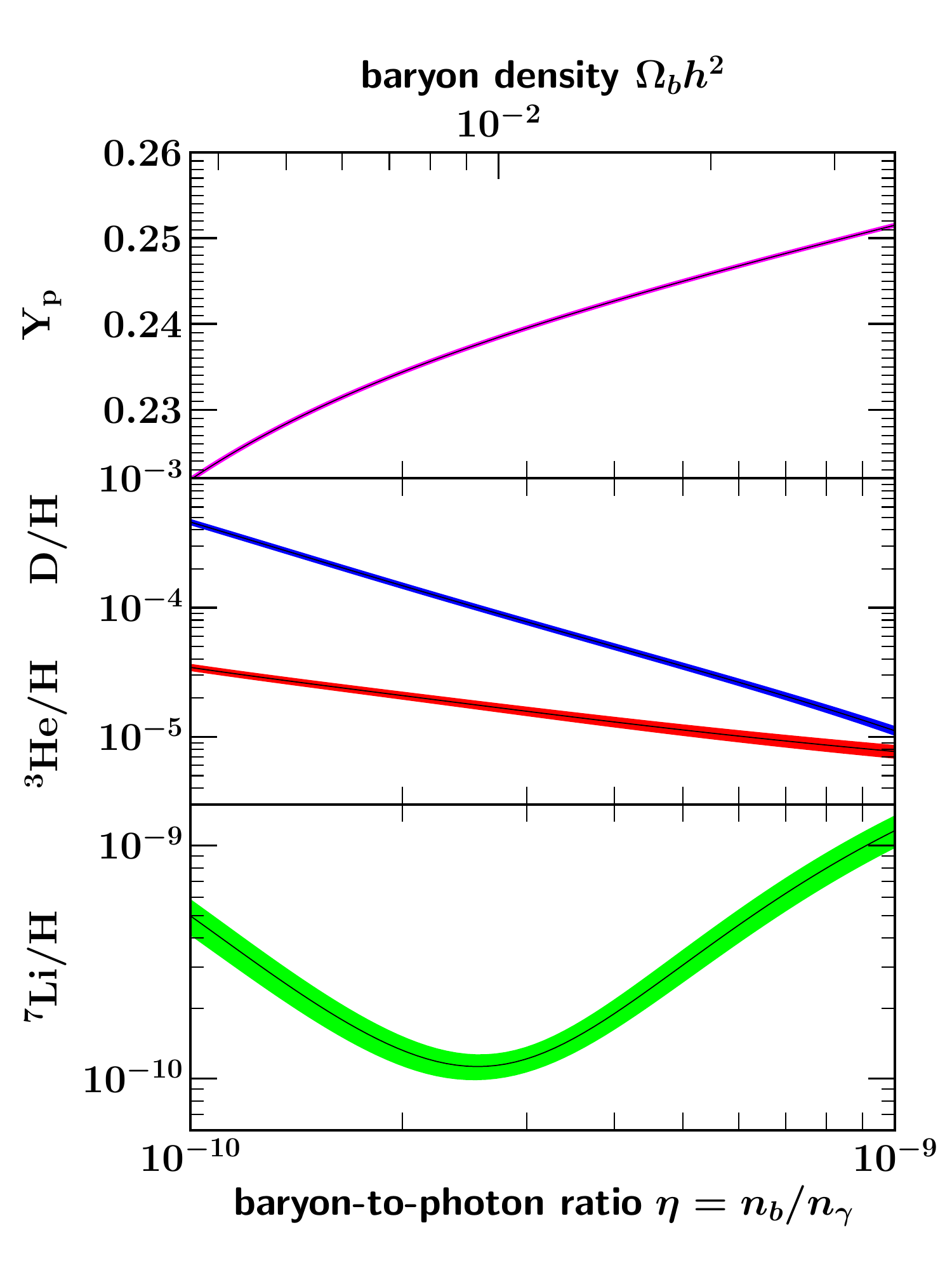}
\caption{
Primordial abundances of the light nuclides as a function of
cosmic baryon content, as predicted by SBBN (``Schramm plot'').
These
results assume $N_\nu = 3$ and the current measurement of the neutron lifetime $\tau_n = 879.4 \pm 0.6$ s. 
Curve widths show $1\sigma$ errors.
\label{fig:schramm}
}
\end{figure}
\end{center}

For $N_\nu = 3$
the baryon-to-photon ratio is the only free
parameter.  Our Monte Carlo error propagation is
summarized in Figure~\ref{fig:schramm}, which plots the light element abundances
as a function of the baryon density (upper scale) and $\eta$ (lower scale).
In the left panel,
the abundance for He is shown as the mass fraction $Y$, while the abundances of
D, \he3, and \li7 are shown as abundances by number relative to H. 
The line through each colored band is our mean value as a function of $\eta$. 

\begin{figure}[!htb]
    \centering
    \includegraphics[width=0.7\textwidth]{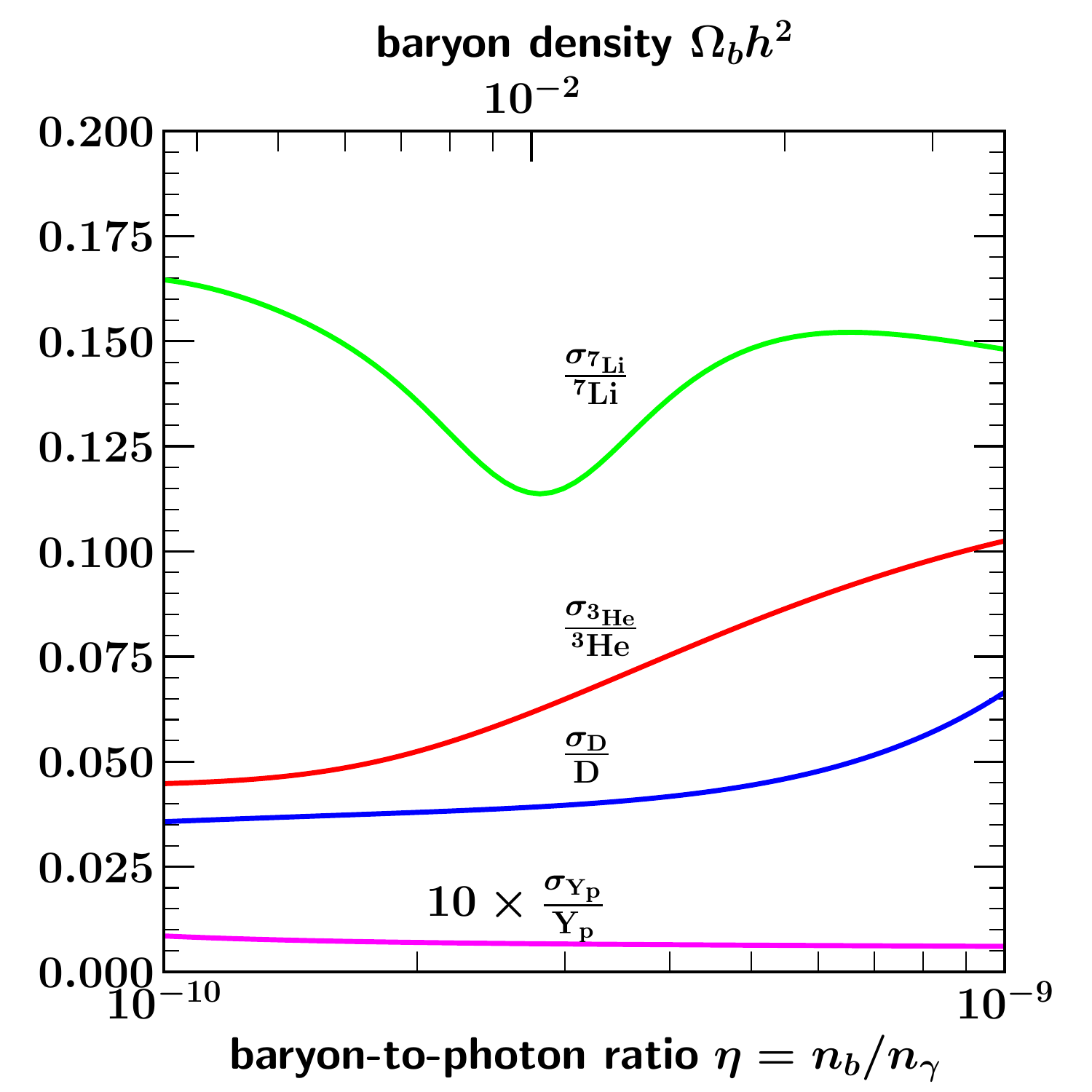}
    \caption{Fractional uncertainties in the light element abundance predictions shown in Fig.~\ref{fig:schramm}.  For each species $i$, we plot ratio of the standard deviation $\sigma_i$ to the mean $\mu_i$, as a function of baryon-to-photon ratio. The relative uncertainty of the \he4 abundance has been multiplied by a factor of 10. }
    \label{fig:schrammerrors}
\end{figure}

The thickness of each curve depicts the $\pm 1 \sigma$ spread in the predicted abundances.  The relative uncertainty,
(the thickness of the curves, relative to the central value) is shown more clearly in Figure~\ref{fig:schrammerrors}. As one can see, the \li7 abundance remains the most uncertain, with a relative uncertainty of approximately 15\%. 
The uncertainty in the deuterium abundance ranges from 3.5-6.5\%,
while the calculated uncertainty in \he4 is only 0.1\% (note that it is amplified by a factor of 10 in the figure for clarity). 

The neutron mean-life of $\tau_n = 879.4 \pm 0.6$ s from Eq. (\ref{nlife})
was used here. The uncertainty of 0.6 s
is one of the dominant sources of uncertainty in the final \he4 abundance (the thickness of the \he4 band). Had we used the previous value of 880.3 s,
the central line would shift upwards by
about the thickness of the central line as shown.

In addition to the dependence on $\eta, \tau_n$, and $N_\nu$, the light elements are sensitive to key nuclear rates. These sensitivities can be quantified as 
\beq
\label{eqn:sens}
X_i = X_{i,0}\prod_n \left(\frac{p_n}{p_{n,0}}\right)^{\alpha_n} \, ,
\eeq
where $X_i$ represents either the helium mass fraction or the abundances of the other
light elements by number. The $p_n$ represent input quantities to the BBN calculations, whose power-law scaling is the log derivative
$\alpha_i = \partial \ln X_i/\partial \ln p_i$. 
Below are the scalings with these inputs, as defined in detail in CFOY. Here, and below, we define $\e10 \equiv 10^{10}\eta$. 
\beqar
Y_p &=& 0.24696\!\left(\frac{\e10}{6.129}\right)^{0.039}\!\!\left(\frac{N_\nu}{3.0}\right)^{0.163}\!\!\left(\frac{G_N}{G_{N,0}}\right)^{0.357}\!\!\left(\frac{\tau_n}{879.4 s}\right)^{0.729} \nonumber \\
 &\times& \left[ p(n,\gamma)d\right]^{0.005}\left[ d(d,n)\he3\right]^{0.006}\left[ d(d,p)t\right]^{0.005} 
\label{yfit}
\eeqar
\beqar
\frac{\rm D}{\rm H} &=& 2.559\!\times\! 10^{-5}\!\left(\frac{\e10}{6.129}\right)^{-1.597}\!\!\left(\frac{N_\nu}{3.0}\right)^{0.396}\!\!\left(\frac{G_N}{G_{N,0}}\right)^{0.952}\!\!\left(\frac{\tau_n}{879.4 s}\right)^{0.409} \nonumber  \\
&\times& \left[ p(n,\gamma)d\right]^{-0.193}\left[ d(d,n)\he3\right]^{-0.529}\left[ d(d,p)t\right]^{-0.470} \nonumber \\
&\times& \left[d(p,\gamma)\he3\right]^{-0.315}\left[\he3(n,p)t\right]^{0.023}\left[\he3(d,p)\he4\right]^{-0.012} 
\eeqar
\beqar
\frac{\he3}{\rm H} &=& 9.965\!\times\! 10^{-6}\!\left(\frac{\e10}{6.129}\right)^{-0.583}\!\!\left(\frac{N_\nu}{3.0}\right)^{0.139}\!\!\left(\frac{G_N}{G_{N,0}}\right)^{0.335}\!\!\left(\frac{\tau_n}{879.4 s}\right)^{0.144} \nonumber \\
&\times& \left[ p(n,\gamma)d\right]^{0.087}\left[ d(d,n)\he3\right]^{0.213}\left[ d(d,p)t\right]^{-0.265} \nonumber  \\
&\times& \left[d(p,\gamma)\he3\right]^{0.376}\left[\he3(n,p)t\right]^{-0.169}\left[\he3(d,p)\he4\right]^{-0.763}\left[t(d,n)\he4\right]^{-0.009} 
\eeqar
\beqar
\frac{\li7}{\rm H} &=& 4.702\!\times\! 10^{-10}\!\left(\frac{\e10}{6.129}\right)^{2.094}\!\!\left(\frac{N_\nu}{3.0}\right)^{-0.280}\!\!\left(\frac{G_N}{G_{N,0}}\right)^{-0.719}\!\!\left(\frac{\tau_n}{879.4 s}\right)^{0.434} \nonumber  \\
&\times& \left[ p(n,\gamma)d\right]^{1.323}\left[ d(d,n)\he3\right]^{0.696}\left[ d(d,p)t\right]^{0.064} \nonumber  \\
&\times& \left[d(p,\gamma)\he3\right]^{0.589}\left[\he3(n,p)t\right]^{-0.267}\left[\he3(d,p)\he4\right]^{-0.754}\left[t(d,n)\he4\right]^{-0.023} \nonumber  \\
&\times& \left[\he3(\alpha,\gamma)\be7\right]^{0.964}\left[\be7(n,p)\li7\right]^{-0.707}\left[\li7(p,\alpha)\he4\right]^{-0.055}\left[t(\alpha,\gamma)\li7\right]^{0.029} \nonumber  \\
&\times&\left[\be7(n,\alpha)\he4\right]^{-0.001}\left[\be7(d,p)\he4\he4\right]^{-0.008}
\label{li7fit}
\eeqar
These sensitivities are summarized as well in Table \ref{tab:sens}. The differences in these sensitivities compared with CFOY are minor.

\begin{table}[!htb]
\caption{Summary of the sensitivities, $\alpha_n$'s defined in Eq.~(\ref{eqn:sens}) for each of the light element abundance predictions, varied with respect to key parameters and reaction rates,
and run for $\eta_{10}=6.129$.
\label{tab:sens}
}
\begin{tabular}{|l|c|c|c|c|}
\hline
Variant &	$Y_p$ &	D/H &	\he3/H & \li7/H \\
\hline
\hline
$\eta$ (6.129$\times 10^{-10}$) &	0.039 &	-1.597 &-0.583 &	2.094  \\
\hline
$N_\nu$ (3.0) & 0.163 & 0.396 & 0.139 & -0.280  \\
\hline
$G_N$&	0.357&	0.952&	0.335&	-0.719 \\
\hline
n-decay&	0.729&	0.409&	0.144&	0.434 \\
\hline
p(n,$\gamma$)d&	0.005&	-0.193&	0.087&	1.323 \\
\hline
\he3(n,p)t&	0.000&	0.023	&-0.169&	-0.267 \\
\hline
\be7(n,p)\li7&	0.000&	0.000&	0.000&	-0.707 \\
\hline
d(p,$\gamma$)\he3&	0.000&	-0.315&	0.376&	0.589 \\
\hline
d(d,$\gamma$)\he4&	0.000&	0.000&	0.000&	0.000 \\
\hline
\li7(p,$\alpha$)\he4&	0.000&	0.000&	0.000&	-0.055 \\
\hline
t($\alpha,\gamma$)\li7&	0.000&	0.000&	0.000&	0.029 \\
\hline
\he3($\alpha,\gamma$)\be7&	0.000&	0.000&	0.000&	0.964 \\
\hline
d(d,n)\he3&	0.006&	-0.529&	0.213&	0.696 \\ 
\hline
d(d,p)t&	0.005&	-0.470&	-0.265&	0.064 \\
\hline
t(d,n)\he4&	0.000&	0.000&	-0.009&	-0.023 \\
\hline
\he3(d,p)\he4&	0.000&	-0.012&	-0.763&	-0.754 \\
\hline
\be7(n,$\alpha$)\he4&	0.000&	0.000&	0.000&	-0.001 \\
\hline
\be7(d,p)$\alpha$ $\alpha$&	0.000&	0.000&	0.000&	-0.008 \\
\hline
\hline
\end{tabular}
\end{table}

We have also tested the effect of certain rates on our baseline results. 
These are given in the appendices. 
In Appendix \ref{sect:7benp-fits}, we consider various measurements of the $\be7(n,p)\li7$ rate, and in Appendix \ref{sect:other-rates-app}, we show the effect of the new $\be7(n,\alpha)\he4$ and $\be7(d,p)\he4$ measurements. The sensitivities to these two destruction rates have been included in Table \ref{tab:sens}.
The low sensitivity in both cases foreshadows our conclusion that these rates indeed are quite subdominant.
In Appendix \ref{sect:dpg-app} we use the theory-based rate for $d(p,\gamma)\he3$.

\section{{\em Planck} Likelihood Functions}
\label{sect:cmbpars}

As we indicated above, we 
begin the {\em Planck} likelihood chains
which are based on temperature and polarization data, TT+TE+EE+lowE,
as well as CMB lensing. We use the chains which do not assume any BBN relation between the helium abundance and baryon density, and we consider separately the case where $N_\nu = 3$ is fixed, and left free. 
We will in addition compare these baseline results to other choices of CMB chains where we drop the inclusion of lensing, and where in addition to lensing we include data from baryon acoustic oscillations (BAO). To further test the robustness of our results, we also consider {\em Planck} chains based on TT+lowE, TE+lowE,
and EE+lowE.  

We are now in a position to combine
the various likelihood functions described in the previous sections. 
These include the CMB likelihood
functions, ${\mathcal L}_{\rm CMB}(\eta,Y_p)$ and ${\mathcal L}_{\rm NCMB}(\eta,Y_p,N_\nu)$ described in \S \ref{sect:CMB}, 
the BBN likelihood functions 
${\mathcal L}_{\rm BBN}(\eta;X)$ and ${\mathcal L}_{\rm NBBN}(\eta,N_\nu;X)$ described in \S \ref{BBNlike}, as well as the observational likelihoods ${\mathcal L}_{\rm OBS}(X)$ which are assumed to be Gaussians based on Eq. (\ref{ypr}) and (\ref{d/h})
for $X \in (Y_p,{\rm D/H})$.

\subsection{2D Fits:  BBN with CMB-Determined Baryon and Helium Abundances}

We first consider models fixing $N_\nu=3$.  In this case standard BBN is a one-parameter theory, depending only on the cosmic baryon density.  Moreover, in the conventional cosmology,
$\eta$ and $Y_p$ do not change between nucleosynthesis and
recombination,
so we may combine the information from these epochs.
The various likelihood functions can be convolved in a number of different ways.  For element abundance determinations, we can compare the 
observational likelihood ${\mathcal L}_{\rm OBS}(X)$ with the following convolution of the CMB and BBN likelihood functions
\beq
{\mathcal L}_{\rm CMB-BBN}(X_i) \propto \int 
  {\mathcal L}_{\rm CMB}(\eta,Y_p) \
  {\mathcal L}_{\rm BBN}(\eta;X_i) \ d\eta \, ,
\label{CMB-BBN}
\eeq
where we normalize each of the likelihood functions so that their peak takes the common value of 1. Thus
we arrive at zero-parameter predictions of abundances for all of the light nuclides.
In the case of \he4, we can also marginalize over $\eta$ to obtain
a CMB-only likelihood function 
\beq
{\mathcal L}_{\rm CMB}(Y_p) \propto \int 
  {\mathcal L}_{\rm CMB}(\eta,Y_p) \ d\eta \, .
  \label{CMByp}
\eeq

\begin{center}
\begin{figure}[!htb]
\includegraphics[width=0.80\textwidth]{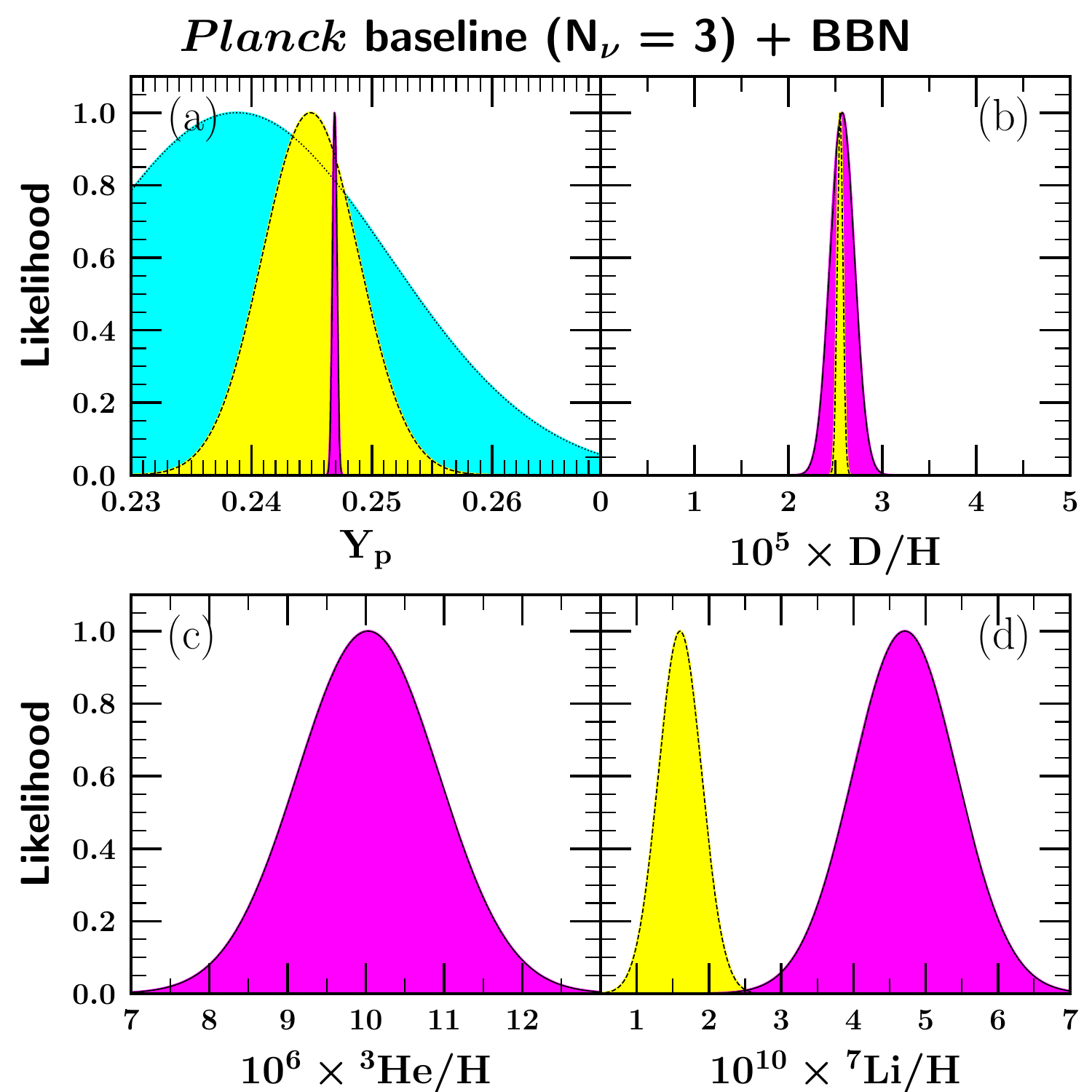}
\caption{
Light element abundance likelihood functions.  Shown are likelihoods for each of the light nuclides,
normalized to show a maximum value of 1.  The solid-lined,
dark-shaded (purple) curves are 
the BBN+CMB predictions, based on {\em Planck} inputs as discussed
in the text.  The dashed-lined, light-shaded 
(yellow) curves show astronomical measurements
of the primordial abundances, for all but \he3 where 
reliable primordial abundance measures do not exist.  For 
\he4, the dotted-lined, medium-shaded (cyan) curve shows
the independent CMB determination of \he4.
We see excellent agreement for D/H, good agreement for \he4, and strong discrepancy in \li7 constitutes the persistent lithium problem.  
\label{fig:2x2abs_2d}
}
\end{figure}
\end{center}

Figure~\ref{fig:2x2abs_2d} shows
the comparison of these likelihood functions for (a) $Y_p$ (upper left), (b) D/H (upper right), (c) \he3/H (lower left), and (d) \li7/H (lower right). In the case of \he4, we show all three likelihood functions. The combined CMB-BBN likelihood from Eq. (\ref{CMB-BBN}), ${\mathcal L}_{\rm CMB-BBN}(Y)$, is shaded purple. The observational likelihood, ${\mathcal L}_{\rm OBS}(Y)$ from Eq. (\ref{ypr})
is shaded yellow. The CMB-only likelihood,
${\mathcal L}_{\rm CMB}(Y_p)$, is shaded cyan. The largest change in these results from {\em Planck} 2015 is seen in the CMB-only result for $Y_P$ which shifted down
from 0.250 to 0.239 with a slightly lower uncertainty of 0.013 compared with 0.14 in 2015. Given the uncertainties in these likelihood distributions,
as seen by the width of the likelihoods, all three are in good agreement.

In the cases of D/H and \li7/H, we are able to compare the observational likelihoods
(shaded yellow) with the combined CMB-BBN likelihoods (shaded purple).
One can see the excellent agreement between the observational value of D/H (in Eq. (\ref{d/h})) and the CMB-BBN predicted value. In contrast, the is a clear mismatch between the observational and CMB-BBN likelihoods for \li7. 

There continue to be two directions of inquiry suggested by the remarkable contrast between the excellent concordance for D and \he4 observations, and the longstanding lithium problem \cite{cfo5}.  One approach is to assume the lithium problem points to new physics at play in the early Universe, pushing us beyond the standard cosmology and standard BBN.  
The other approach to assume the lithium problem will find its solution in observational or astrophysical systematics.  For example, 
internal stellar depletion may be important, making the observations of stellar Li
non-representative of their initial
and near primordial abundance.
In this scenario we retain the standard cosmology, ignore the lithium data, and concentrate on \he4 and D/H to probe the cosmic baryon density.  

Finally, we show only the CMB-BBN likelihood for \he3,
because of the lack of a reliable method of extracting a {\em primordial} abundance from existing \he3 observational data.

The CMB-BBN likelihoods in Fig.~\ref{fig:2x2abs_2d}
are summarized by
the predicted abundances 
\beqar
Y_p &=& 0.24691\pm0.00018 \qquad (0.24691) \label{meyp} \\
{\rm D/H} &=& (2.57\pm0.13)\times 10^{-5} \qquad (2.57 \times 10^{-5})\label{medh}\\
\he{3}/{\rm H} &=& (10.03\pm 0.90)\times 10^{-6} \qquad (10.03 \times 10^{-6}) \label{mehh} \\
\li{7}/{\rm H} &=& (4.72\pm 0.72)\times 10^{-10} \qquad (4.71 \times 10^{-10}) \label{melh}
\eeqar
where the central values give the mean,
and the error gives the $1\sigma$ variance.
The final number in parentheses gives the value at the peak of the distribution.

We compare our results to previous results in CFOY \cite{CFOY}
and ref. \cite{coc18} in Table \ref{tab:re}.
The values in Eqs.~(\ref{meyp})-(\ref{melh}) differ slightly from those given in
Table \ref{tab:re} as the latter were evaluated using central values of all inputs at a single value of $\eta_{10} = 6.129$. 

\begin{table}[ht]
\caption{Comparison of BBN Results
\label{tab:re}
}
\begin{tabular}{|l|c|c|c|c|c|c|c|}
\hline
 & $\eta_{10}$ & $N_{\nu}$ & $Y_p$ & D/H & \he3/H & \li7/H  \\
\hline
Ref. \cite{CFOY}& 6.10 & 3 & 0.2470 & 2.579 $\times$ $10^{-5}$& 0.9996 $\times$ $10^{-5}$ &  $4.648 \times$ $10^{-10}$  \\
\hline
Ref. \cite{coc18}& 6.091 & 3 & 0.2471 & 2.459 $\times$ $10^{-5}$& 1.074 $\times$ $10^{-5}$ &  $5.624 \times$ $10^{-10}$  \\
\hline
this work & 6.129 & 3 & 0.2470 & 2.559 $\times$ $10^{-5}$& 0.9965 $\times$ $10^{-5}$ &  $4.702 \times$ $10^{-10}$  \\
\hline
\end{tabular}
\end{table}

\begin{figure}[!htb]
    \centering
    \includegraphics[width=0.75\textwidth]{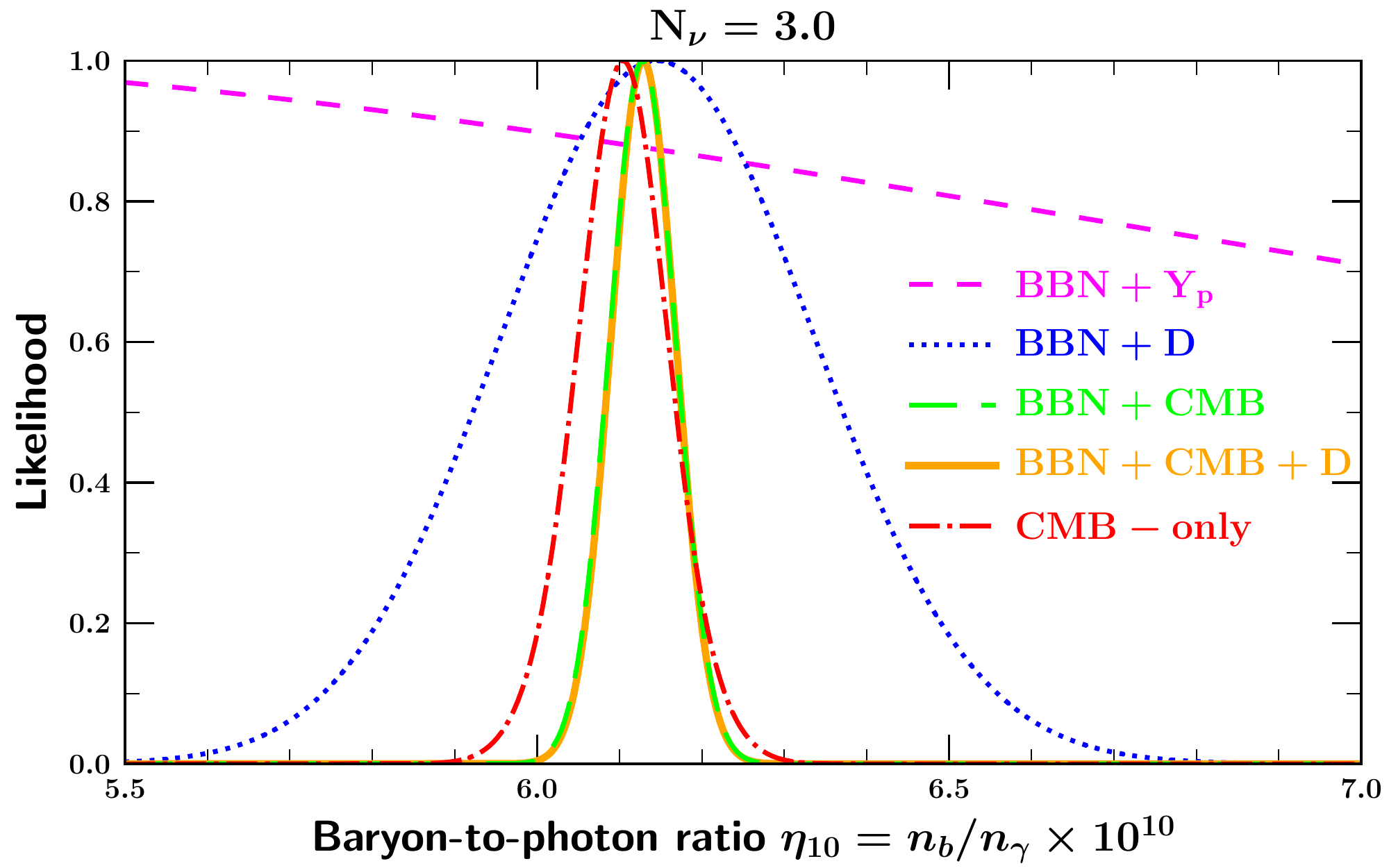}
    \caption{Baryon-to-photon ratio determinations for different combinations of light element and CMB data.  We show BBN-only predictions based on D (dotted purple) and $Y_p$ (dashed magenta), and CMB-only predictions in dot-dashed red.  BBN+CMB (green dot-long dashed) uses the {\em Planck} $Y_p$ data, while the tightest combined constraints, BBN+CMB+D (solid orange) further include the observed D/H.}
    \label{fig:SBBN-eta-baseline}
\end{figure}

There are additional ways of integrating over our various likelihood functions. We can for example, simply marginalize the CMB likelihood function
 over $Y_P$ to obtain
a CMB-only likelihood function
of $\eta$
\beq
{\mathcal L}_{\rm CMB}(\eta) \propto \int 
  {\mathcal L}_{\rm CMB}(\eta,Y_p) \ dY_p \, .
\eeq
This is plotted in Fig.~\ref{fig:SBBN-eta-baseline}
as the red dot-dashed curve. Its mean and standard deviation are given in Table \ref{tab:eta}. 
Also given in Table \ref{tab:eta} is the position of the peak of the distribution. Its difference from the mean value is a measure of the mode skewness of the distribution. It is always very small. 

\begin{table}[!htb]
\caption{Constraints on the baryon-to-photon ratio, using different combinations of observational constraints.  We have marginalized over $Y_p$ to create 1D $\eta$ likelihood distributions. Given are both the mean (and its uncertainty)
as well as the value of $\eta$ at the peak of the distribution.
\label{tab:eta}
}
\vskip .2in
\begin{tabular}{|l|c|c|}
\hline
 Constraints Used & mean $\e10$ & peak $\e10$ \\
\hline
CMB-only & $6.104\pm 0.058$ & 6.104 \\
\hline
\hline
BBN+$Y_p$ & $6.741 {}^{+1.220}_{-3.524}$ & 4.920 \\
\hline
BBN+D & $6.148\pm 0.191$ & 6.145 \\
\hline
BBN+$Y_p$+D & $6.143\pm 0.190$ & 6.140 \\
\hline
CMB+BBN & $6.128\pm 0.040$ & 6.128 \\
\hline
CMB+BBN+$Y_p$ & $6.128\pm 0.040$ & 6.128 \\
\hline
CMB+BBN+D & $6.129\pm 0.039$ & 6.129 \\
\hline
\hline
CMB+BBN+$Y_p$+D & $6.129\pm 0.039$ & 6.129 \\
\hline
\end{tabular}
\end{table}

The likelihood function ${\mathcal L}_{\rm CMB}(\eta)$ uses no information from BBN.
In particular it does not use the BBN relation between $\eta$ and $Y_p$. This relation can be folded in by computing
\beq
{\mathcal L}_{\rm CMB-BBN}(\eta) \propto \int 
  {\mathcal L}_{\rm CMB}(\eta,Y_p) \
  {\mathcal L}_{\rm BBN}(\eta;Y_p) \ dY_p \, ,
  \label{CMB-BBN-eta}
\eeq
which is shown in Fig.~\ref{fig:SBBN-eta-baseline} by the green dot-long dashed curve.

As is well known and seen in Fig.~\ref{fig:schramm}, there is a weak dependence of $Y_p$ on $\eta$. As a result, though one can form a likelihood
function from BBN and $Y_p$ alone,
\beq
{\mathcal L}_{\rm BBN-OBS}(\eta) \propto \int 
  {\mathcal L}_{\rm BBN}(\eta;X_i) \
  {\mathcal L}_{\rm OBS}(X_i) \ dX_i \, ,
  \label{OBS-eta}
\eeq
with $X_i = Y_p$, it is not very instructive. It is
shown in Fig.~\ref{fig:SBBN-eta-baseline}
by the very broad magenta dashed curve.
In contrast, D/H is a very good baryometer,
and substituting $X_i = \rm D/H$ in Eq. (\ref{OBS-eta}) yields the purple dotted curve in Fig.~\ref{fig:SBBN-eta-baseline}.
Finally, we can convolve all three primary
likelihood functions as
\beq
{\mathcal L}_{\rm CMB-BBN-OBS}(\eta) \propto \int 
{\mathcal L}_{\rm CMB}(\eta,Y_p)
  {\mathcal L}_{\rm BBN}(\eta;X_i) \
  {\mathcal L}_{\rm OBS}(X_i) \ \prod_i dX_i \, ,
  \label{CMB-OBS-eta}
\eeq
which is shown as the solid orange curve in Fig.~\ref{fig:SBBN-eta-baseline}.
With the exception of ${\mathcal L}_{\rm BBN-OBS}(\eta)$
using $Y_p$, which carries little information, all of the
likelihoods are remarkably consistent which is another reflection of the agreement between the BBN prediction of D/H at the CMB-determined value of $\eta$ and the observationally-determined value of D/H. 
The mean, standard deviations, and peaks of all of these likelihood functions are summarized in Table \ref{tab:eta}.

\begin{center}
\begin{figure}[!htb]
\includegraphics[width=0.75\textwidth]{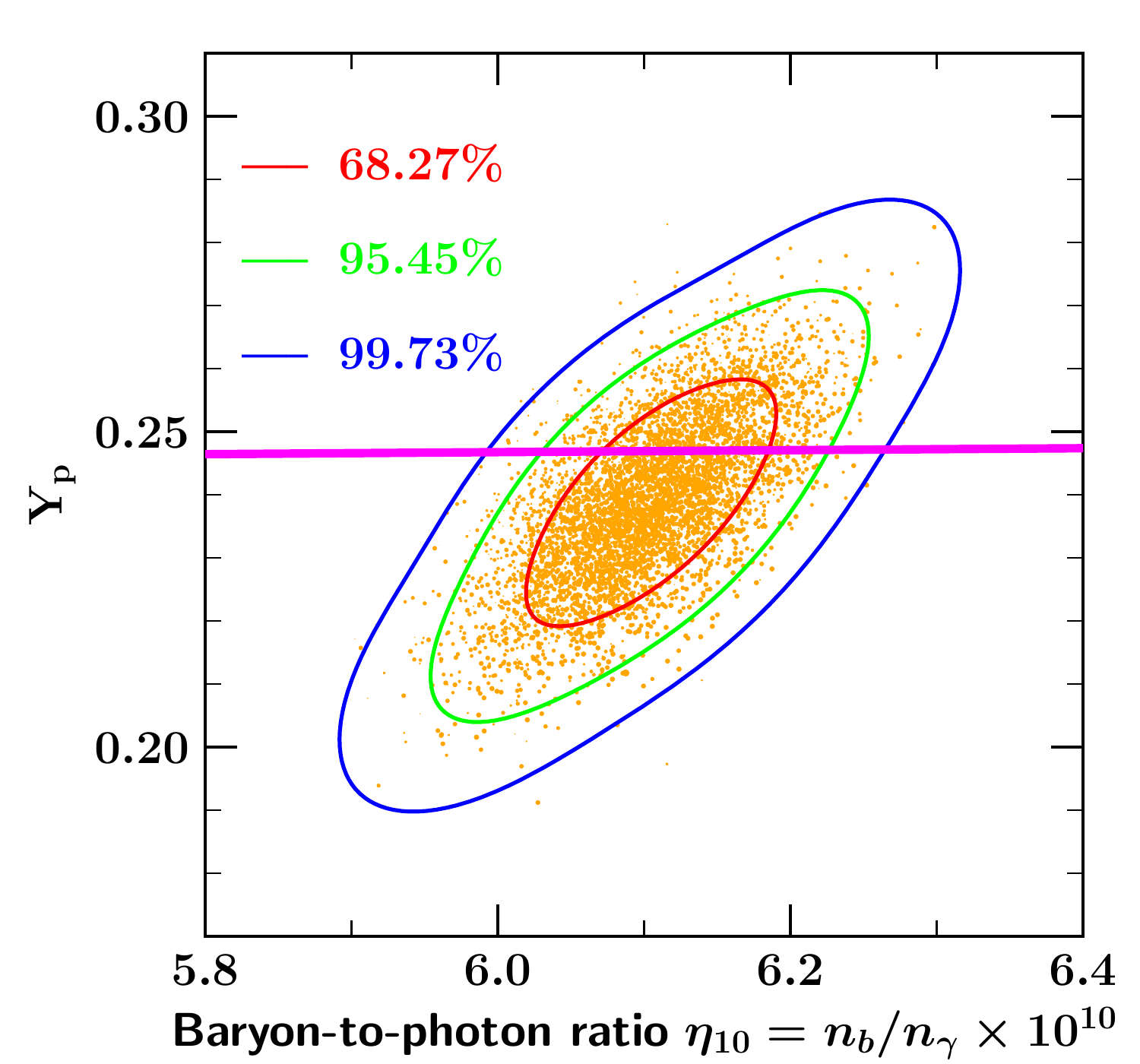}
\caption{
The 2D likelihood function contours derived ${\mathcal L}_{\rm CMB}(\eta,Y_p)$.  
The 3-$\sigma$ BBN prediction for the helium mass fraction is shown by the nearly horizontal colored band.
\label{fig:L2D1}
}
\end{figure}
\end{center}

As one can see from Table \ref{tab:eta}, the BBN + D likelihood gives  $\eta_{10} \simeq 6.15$ and is slightly lower than that found in CFOY ($\eta_{10} \simeq 6.18$). This is primarily due to the very slight shift in the observational value of D/H used. With
all other factors fixed the change in $\eta$ can be estimated from the sensitivities discussed earlier and we expect $\delta {\rm (D/H)/(D/H)} \simeq -1.6 \, \delta \eta/\eta$. In contrast, all of the CMB+BBN determinations of $\eta_{10}$ are increased from 
$\simeq 6.10$ (in 2015) to roughly 6.13 presently. This tendency can be understood using Fig.~\ref{fig:L2D1} 
which shows contours of the 2-D likelihood ${\mathcal L}_{\rm CMB}(\eta,Y_p)$ for fixed $N_\nu = 3$.
Also shown is the BBN relation for $Y_p(\eta)$ which appears
as a nearly horizontal line over this range in $\eta$.
Thus small changes in $\eta$ barely affect the peak of the
likelihood function (shaded purple) in Fig.~\ref{fig:2x2abs_2d}.
In contrast, 
the CMB contours show a significantly stronger and positive correlation between
the CMB-only determined baryon density and helium abundance.
Now, as noted above, one of the more noticeable changes between
{\em Planck} 2015 and 2018 was the CMB-only determination of $Y_p$.
Using {\em Planck} 2015, the peak of the CMB-only distribution of $Y_p$ was high compared to the observational peak and as a result
$\eta$ was found to be lower when BBN was included (relative to the CMB-only value of $\eta$). Currently, as one sees in Fig.~\ref{fig:2x2abs_2d}, the CMB-only distribution for $Y_p$
sits below the observational value and as a result
requires a higher value of $\eta$ when the distributions are convolved. This is precisely what we find. Our final combined value for the baryon-to-photon ratio, is therefore 
\beq
\eta = (6.129 \pm 0.039) \times 10^{-10} \ , \qquad \omb = 0.02239 \pm 0.00014 \, .
\eeq

\begin{center}
\begin{figure}[htb]
\includegraphics[width=0.75\textwidth]{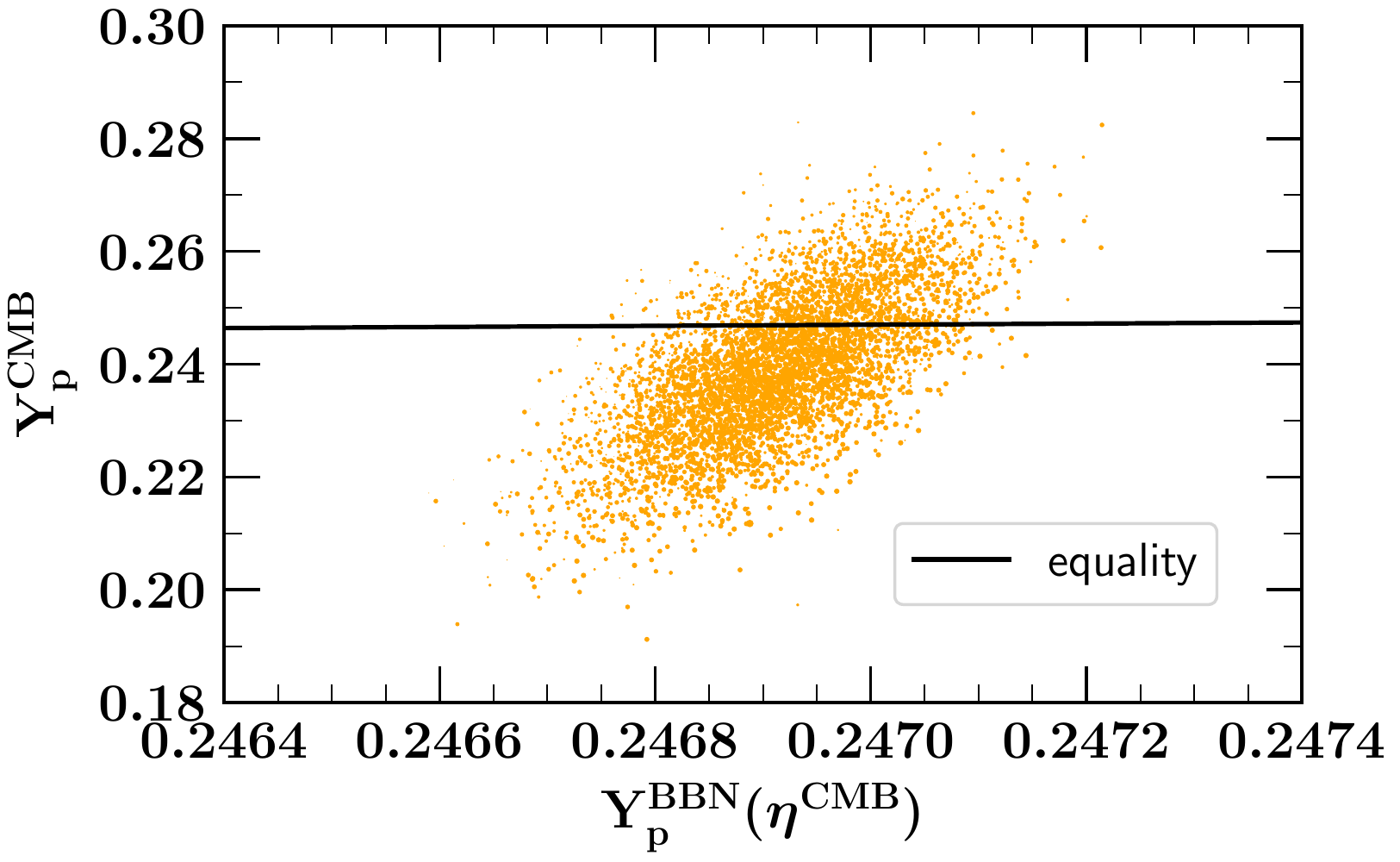}
  \caption{
    An assessment of the status of \he4.
    The points compare the primordial \he4 abundance directly from the CMB
    damping tail $Y_{\rm cmb}$,
    with \he4 as calculated from the CMB $\eta$ and BBN theory:
    $Y_{\rm bbn}(\eta_{\rm cmb})$.  Note that the vertical BBN scale is much tighter than the horizontal CMB scale.  The solid line gives
    the expected $Y_{\rm cmb} = Y_{\rm bbn}(\eta_{\rm cmb})$ trend, which appears almost horizontal within these 
    very different axis scales.
  }
  \label{2dyp}
\end{figure}
\end{center}

We can also plot the 2d CMB likelihood function, ${\mathcal L}_{\rm CMB}(\eta,Y_p)$ showing instead of $\eta$, the BBN value of $Y_p$ at that value of $\eta$. This is shown in Fig.~\ref{2dyp}. That is, 
we use the peak of ${\mathcal L}_{\rm BBN}(\eta,X_i)$ to convert
$\eta$ to $Y_p^{\rm BBN}$. The solid line shows
the equality between the two values of $Y_p$
appears nearly horizontal because of the very different
scales used on each axis.

\subsection{3D fits: Variable $N_\nu$}

\begin{center}
\begin{figure}[!htb]
\includegraphics[width=0.75\textwidth]{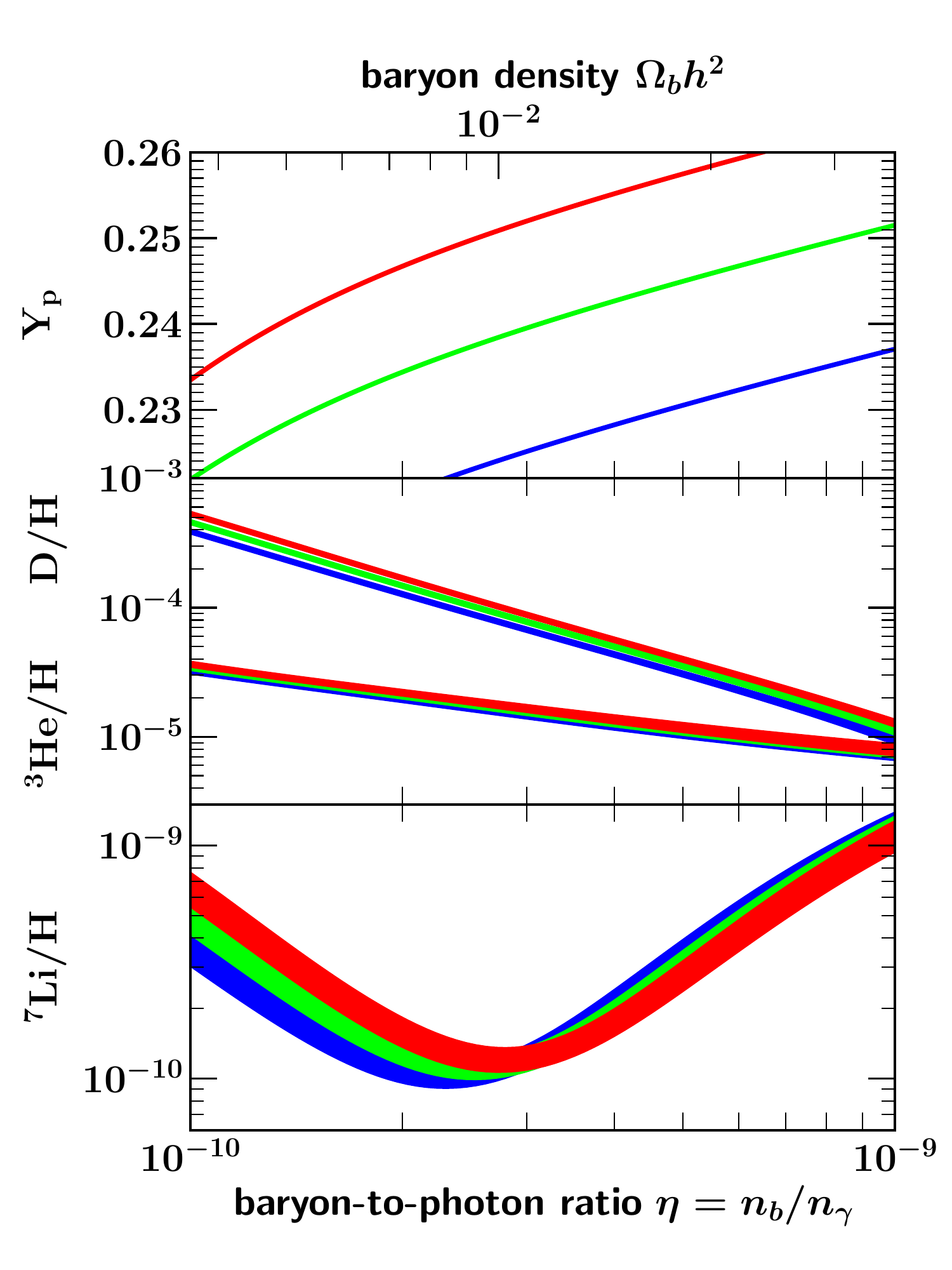}
\caption{
The sensitivity of the light element predictions to the number of neutrino species, similar to Figure~\ref{fig:schramm}.  Here, abundances shown by blue, green, and red bands correspond to 
calculated abundances assuming $N_\nu = 2, 3$ and 4 respectively.
\label{fig:SchrammNnu}
}
\end{figure}
\end{center}

We turn now to the set of likelihood functions for NBBN, i.e.,
with
$N_\nu$ allowed to remain free. 
We again combine BBN and CMB information, which implies
that all of $(\eta,Y_p,N_\nu)$ are the same at these
two epochs.
We start by updating the Schramm plot in Fig.~\ref{fig:SchrammNnu}  showing 
the light element abundances as a function of $\eta$.
This figure is actually still based on the 2d likelihood function ${\mathcal L}_{\rm BBN}(\eta;X_i)$ for different
fixed inputs of $N_\nu = 2,3$, and 4. 
As is well known,
increasing the number of neutrino flavors leads to an increase in the expansion rate of the Universe. As a result,  the weak interactions
freeze-out at a higher temperature, $T_f$, and more neutrons are present when BBN begins yielding a higher
value for $Y_p$ \cite{nnu}.
Given an upper bound on $Y_p$ together with a lower bound on $\eta$ \cite{ossty}, one can establish an upper bound to the number of neutrinos \cite{ssg,GN}.  The increase  of the helium mass fraction with $N_\nu$ can clearly be seen in this figure.

As one can see in Fig.~\ref{fig:SchrammNnu}, without a lower bound on $\eta$, it is not possible to set a meaningful upper limit to $N_\nu$, even with a firm upper bound to $Y_p$. 
Prior to the CMB determination of $\eta$, a lower bound
on $\eta$ was inferred from observations of
of D and \he3 implying $N_\nu < 4$ \cite{ytsso,ytsso1}. More rigorous bounds on $N_\nu$ became became possible 
when likelihood techniques were introduced \cite{kk,ot,lisi,bbnt,cfo2,cfos,ms}.

Note that there is a non-negligible dependence of D/H on $N_\nu$ \cite{cfo2}. This is particularly important since
the deuterium abundance is measured much more accurately
than $Y_p$. As we will see, the constraints on $N_\nu$
will be affected by both $Y_p$ and D/H.

\begin{center}
\begin{figure}[!htb]
\includegraphics[width=0.80\textwidth]{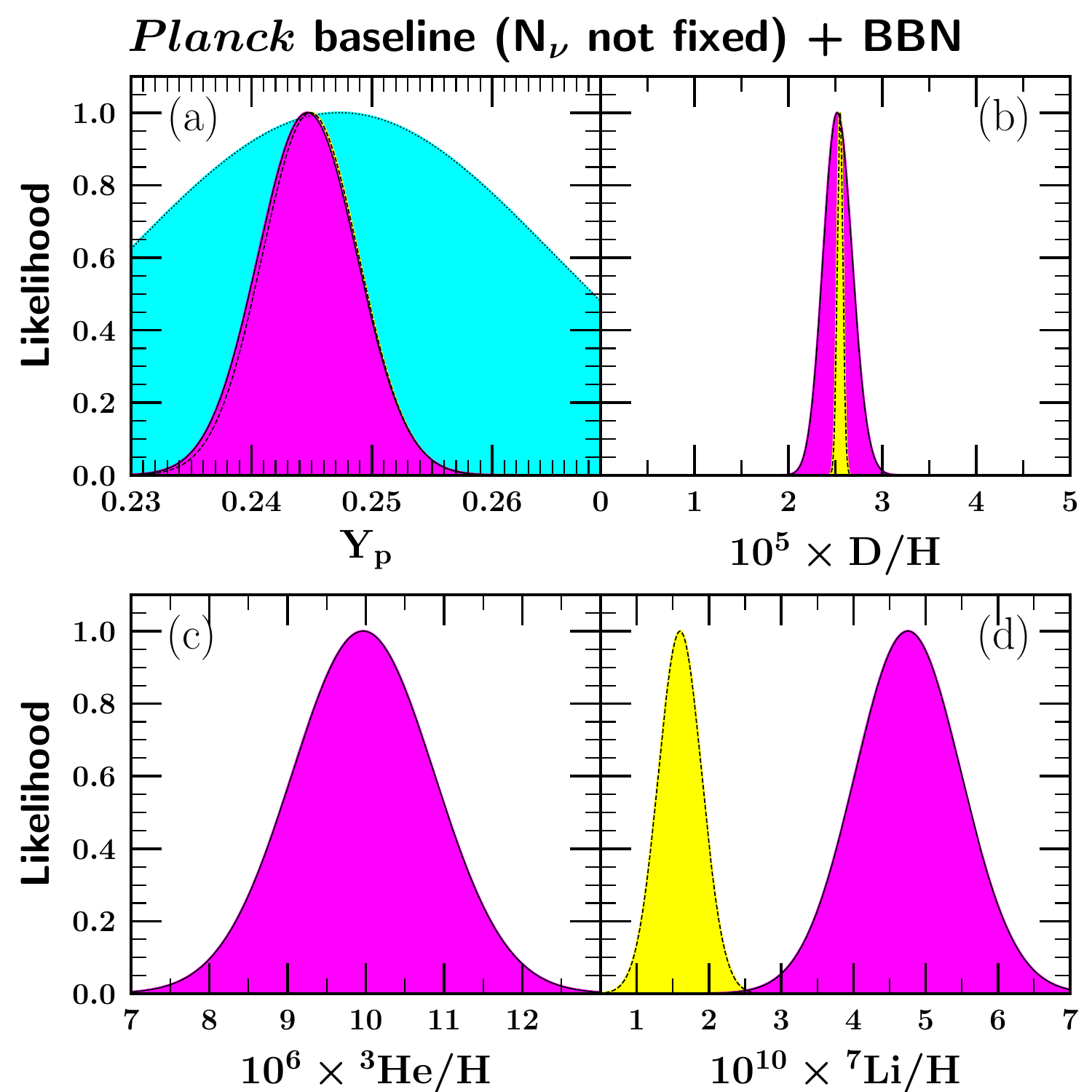}
\caption{ As in Fig.~\ref{fig:2x2abs_2d}, the
light element predictions using the CMB determination of the 
cosmic baryon density when $N_\nu$ is not fixed.  Note that the observational likelihood (in yellow) for \he4 lies almost exactly underneath the CMB-BBN prediction shown in purple. 
}
\label{fig:2x2abs_3d}
\end{figure}
\end{center}

As in the two-dimensional case, we can form
a one-dimensional likelihood function of the element abundances 
using ${\mathcal L}_{\rm NCMB}$
and ${\mathcal L}_{\rm NBBN}$
in place of ${\mathcal L}_{\rm CMB}$ and ${\mathcal L}_{\rm BBN}$
as in Eqs. (\ref{CMB-BBN}) and (\ref{CMByp}). These are shown in Fig.~\ref{fig:2x2abs_3d}. The only 
noticeable change in the likelihood function appears to be ${\mathcal L}_{\rm NBBN}(Y_p)$ which is significantly broader than the case when $N_\nu = 3$ is held fixed.
In this case there is an almost perfect overlap between the CMB and BBN likelihoods.

The resulting mean values, their uncertainties (and peak values)  when
$N_\nu$ is not held fixed are
\beqar
Y_p &=& 0.24465 \pm0.00410 \label{ypnnu} \qquad (0.24498) \\
{\rm D/H} &=& (2.53\pm0.15)\times 10^{-5} \qquad (2.52 \times 10^{-5}) \\
\he{3}/{\rm H} &=& (9.97\pm0.91)\times 10^{-6} \qquad (9.96 \times 10^{-6}) \\
\li{7}/{\rm H} &=& (4.76\pm 0.74)\times 10^{-10} \qquad (4.75 \times 10^{-10}) \, .
\eeqar
Note that the value for $Y_p$ found from the BBN likelihood (\ref{ypnnu})
is almost identical to the observational value in Eq. (\ref{ypr}) which is why the observational likelihood in Fig.~\ref{fig:2x2abs_3d} appears masked. 

\begin{figure}[!htb]
    \centering
    \includegraphics[width=0.95\textwidth]{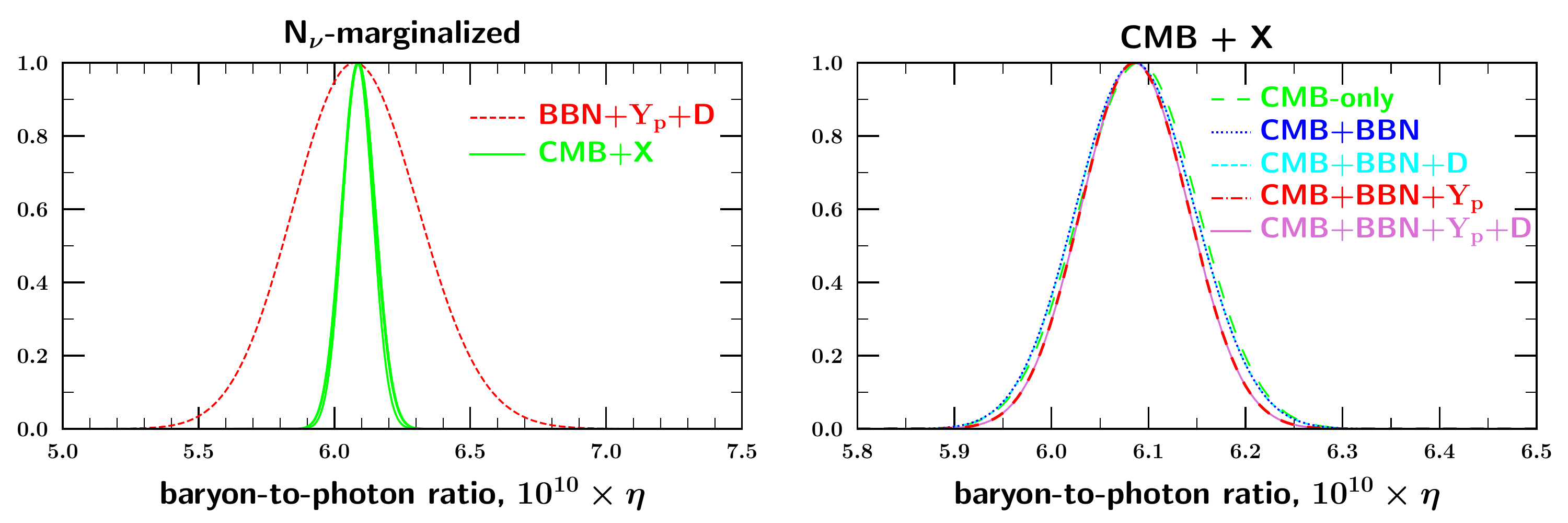}
    \caption{As in Fig.~\ref{fig:SBBN-eta-baseline}, but where $N_\nu$ is allowed to vary, and results marginalized over it.  Right panel is a zoom onto the left panel to illustrate the small changes depending on the light element data included.}
    \label{fig:NBBN-eta-baseline}
\end{figure}

We can also form
a one-dimensional likelihood function of $\eta$
with one of several combinations of 
${\mathcal L}_{\rm NCMB}(\eta,Y_p,N_\nu)$,
${\mathcal L}_{\rm NBBN}(\eta,N_\nu;X_i)$, and ${\mathcal L}_{\rm OBS}(X_i)$.
For example, by integrating over both $Y_p$
and $N_\nu$, using only CMB data, we have 
\beq
{\mathcal L}_{\rm NCMB}(\eta) \propto \int 
{\mathcal L}_{\rm NCMB}(\eta,Y_p,N_\nu)
  \ dY_p \ dN_\nu \, ,
  \label{NCMB-eta}
\eeq
which is shown as the green dashed curved in the right panel of Fig.~\ref{fig:NBBN-eta-baseline}.
Similarly, if we fold in the relation
between $\eta$ and $Y_p$ we have
\beq
{\mathcal L}_{\rm NCMB-NBBN}(\eta) \propto \int 
{\mathcal L}_{\rm NCMB}(\eta,Y_p,N_\nu)
  {\mathcal L}_{\rm NBBN}(\eta, N_\nu;X_i) \
  dY_p dN_\nu \, ,
  \label{NCMB-NBBN-eta}
\eeq
which is shown by the purple dotted curve in the right panel of Fig.~\ref{fig:NBBN-eta-baseline}. We can also fold in the observations of either 
\he4, D/H or both using 
\beq
{\mathcal L}_{\rm NCMB-NBBN-OBS}(\eta) \propto \int 
{\mathcal L}_{\rm NCMB}(\eta,Y_p, N_\nu)
  {\mathcal L}_{\rm NBBN}(\eta,N_\nu;X_i) \
  {\mathcal L}_{\rm OBS}(X_i) \ \prod_i dX_i dN_\nu \, ,
  \label{NCMB-NBBN-NOBS-eta}
\eeq
which depending on the choice of observations is shown by the short dashed cyan curve (using D/H), the red dot-dashed curve (using $Y_p$) or the pink solid curve (using both) in the right panel of Fig.~\ref{fig:NBBN-eta-baseline}. These are collectively shown in the left panel of the same figure by the solid green curve labelled CMB+X. 
If we drop the CMB entirely,
we can write
\beq
{\mathcal L}_{\rm NBBN-OBS}(\eta) \propto \int 
  {\mathcal L}_{\rm NBBN}(\eta,N_\nu;X_i) \
  {\mathcal L}_{\rm OBS}(X_i) \ \prod_i dX_i dN_\nu \, ,
  \label{BBN-OBS-eta}
\eeq
shown by the red short dashed curve in the left panel of Fig.~\ref{fig:NBBN-eta-baseline}. A comparison of the two curves in the left panels shows the strength in determining $\eta$ using the CMB relative to BBN (D/H).

\begin{figure}[!htb]
    \centering
    \includegraphics[width=0.95\textwidth]{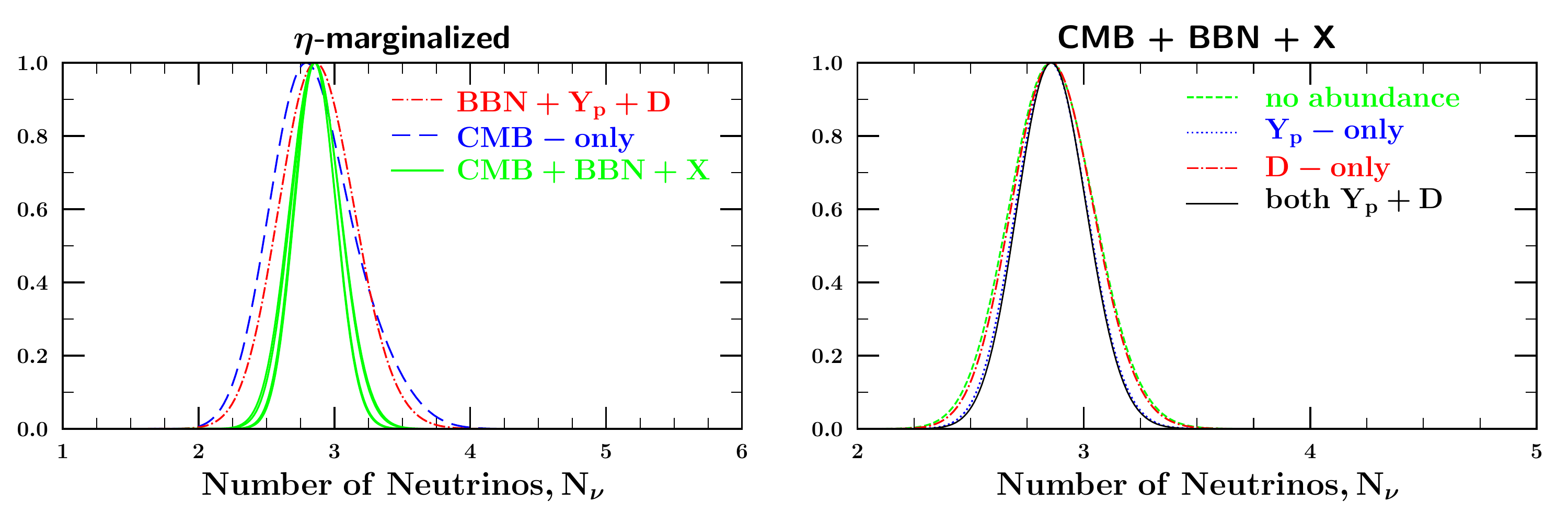}
    \caption{One-dimensional likelihood functions of $N_\nu$.}
    \label{fig:NBBN-nnu-baseline}
\end{figure}

Similarly, we can form one-dimensional likelihood functions of $N_\nu$. For example, 
using the CMB-only likelihood function, we can integrate over $\eta$ and $Y_p$
\beq
{\mathcal L}_{\rm NCMB}(N_\nu) \propto \int 
  {\mathcal L}_{\rm NCMB}(\eta,N_\nu,Y_p)d Y_p d\eta \, ,
  \label{CMB-nnu}
\eeq
which is shown by the blue dashed curve in the left panel of Fig.~\ref{fig:NBBN-nnu-baseline}.
A slightly narrower distribution is found from the BBN convolution with observations 
\beq
{\mathcal L}_{\rm NBBN-OBS}(N_\nu) \propto \int 
  {\mathcal L}_{\rm NBBN}(\eta,N_\nu;X_i) \
  {\mathcal L}_{\rm OBS}(X_i) \ \prod_i dX_i d\eta \, ,
  \label{BBN-OBS-nnu}
\eeq
which is shown by the red dot-dashed curve in the left panel of Fig.~\ref{fig:NBBN-nnu-baseline}.
We can also combine the CMB and BBN with or with out observations
\beq
{\mathcal L}_{\rm NCMB-NBBN-OBS}(N_\nu) \propto \int 
{\mathcal L}_{\rm NCMB}(\eta,N_\nu,Y_p)
  {\mathcal L}_{\rm NBBN}(\eta,N_\nu;X_i) \
  {\mathcal L}_{\rm OBS}(X_i) \ \prod_i dX_i d\eta \, ,
  \label{CMB-BBN-OBS-nnu}
\eeq
which are shown by the solid green curves in the left panel and isolated in the right panel of Fig.~\ref{fig:NBBN-nnu-baseline}. Note that Eq. (\ref{CMB-BBN-OBS-nnu}) is somewhat schematic and must be adjusted according to which observations (if any) are included in the right panel. 

\begin{figure}[!htb]
    \centering
    \includegraphics[width=0.75\textwidth]{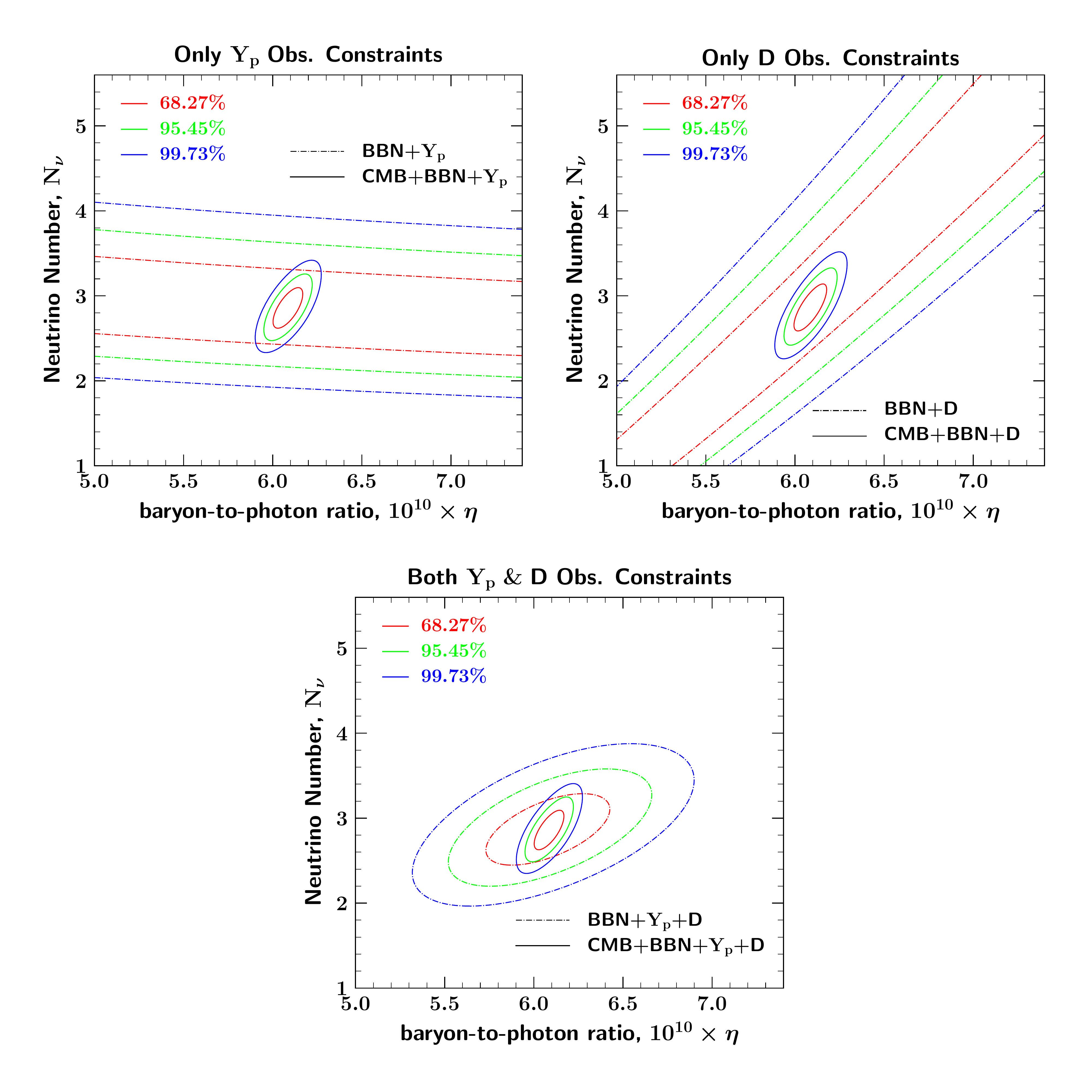}
    \caption{Likelihood distributions in $(N_\nu,\eta)$ space.  Smaller solid contours are the $(68\%,95\%,99\%)$ CL results combining BBN, CMB, light element data.  Larger dotted contours are for BBN and light element data only. Panels use observed data for:  (a) $Y_p$, (b) D/H, and (c) $Y_p$ and D/H.
    We see that BBN-only constraints are weak with only one light element, but give bounds on both $\eta$ and $N_\nu$ when two light elements are used.  Adding the CMB considerably tightens both limits.}
    \label{fig:2deta}
\end{figure}

It is also instructive to consider
two-dimensional likelihood when $N_\nu$
is free. These are shown in Fig.~\ref{fig:2deta} where we consider two convolutions
\beq
{\mathcal L}_{\rm NBBN-OBS}(\eta,N_\nu) \propto \int 
  {\mathcal L}_{\rm NBBN}(\eta,N_\nu;X_i) \
  {\mathcal L}_{\rm OBS}(X_i) \ \prod_i dX_i \, ,
  \label{NBBN-OBS-eta-nnu}
\eeq
and
\beq
{\mathcal L}_{\rm NCMB-NBBN-OBS}(\eta,N_\nu) \propto \int 
{\mathcal L}_{\rm NCMB}(\eta,Y_p)
  {\mathcal L}_{\rm NBBN}(\eta,N_\nu;X_i) \
  {\mathcal L}_{\rm OBS}(X_i) \ \prod_i dX_i \, .
  \label{NCMB-NBBN-OBS-eta-nnu}
\eeq
The first of these likelihoods is used to produce the set of dashed curves in each panel of Fig.~\ref{fig:2deta}, using either
\he4, D/H or both. In each case we show the contours corresponding to the 68.27\% 95.45\% and 99.73\% values of the likelihood function. Once again we see that when the observations of \he4 are used alone,
we get very little information on $\eta$ (over the range shown), but relatively strong limits on $N_\nu$. When observations of D/H are used alone, we get limits on both
$\eta$ and $N_\nu$, though there is a degeneracy and the contours do not close.
They do close when both $Y_P$ and D/H are used as seen in the 3rd panel of Fig.~\ref{fig:2deta}.

Using Eq.~\ref{NCMB-NBBN-OBS-eta-nnu},
we obtain the solid contours in Fig.~\ref{fig:2deta}, all of which are closed. 
As one can see, each of these give relatively strong constraints on both
$\eta$ and $N_\nu$.
The mean and standard deviation of the 
two-dimensional likelihoods, ${\mathcal L}(\eta,N_\nu)$ are summarized in Table \ref{tab:etannu}. Also given is the position of the peak of the likelihood. 

Comparing Table \ref{tab:etannu} to Table \ref{tab:eta}, we see that there is a systematic downward shift in $\eta_{10}$ when
$N_\nu$ is allowed to vary. This can be understood from Figs.~\ref{fig:SchrammNnu} and \ref{fig:2x2abs_3d} where concordance 
is more easily achieved at $N_\nu$ slightly below 3 and 
at slightly lower $\eta_{10}$. The {\em Planck} CMB data alone
also prefer a value of $N_\nu$ slightly below 3, but
with 68\% and 95\% confidence level upper limits on $N_\nu - 3$ of 0.173 and 0.487. When BBN and CMB and observations are combined
we find 
\beq
N_\nu = 2.862 \pm 0.153
\eeq
giving 68\% and 95\% confidence level upper limits of $N_\nu -3 < 0.015$ and 0.168 respectively. 

\begin{table}[!htb]
\caption{The marginalized most-likely values and central 68.3\% confidence limits on the baryon-to-photon ratio and effective number of neutrinos, using different combinations of observational constraints. 
\label{tab:etannu}
}
\vskip.1in
\begin{tabular}{|l|c|c|c|c|}
\hline
 Constraints Used & mean $\eta_{10}$ & peak $\eta_{10}$ & mean $N_\nu$ & peak $N_\nu$ \\
\hline
CMB-only & $6.090\pm0.060$ & 6.090 & $2.859 \pm 0.314$ & 2.793 \\
\hline
\hline
BBN+$Y_p$+D & $6.084 \pm 0.230$  & 6.075 &  $2.878\pm0.278$ & 2.861 \\
\hline
CMB+BBN & $6.087\pm 0.060$ & 6.086 & $2.862 \pm 0.189$ & 2.854 \\
\hline
CMB+BBN+$Y_p$ & $6.086 \pm 0.055$ & 6.085 &  $2.859 \pm 0.157$ & 2.854 \\
\hline
CMB+BBN+D & $6.087 \pm 0.060$ & 6.086 & $2.865 \pm 0.182$ & 2.858 \\
\hline
\hline
CMB+BBN+$Y_p$+D & $6.086 \pm 0.055$ & 6.085 & $2.862 \pm 0.153$ & 2.856 \\
\hline
\end{tabular}
\end{table}

\subsection{Other {\em Planck} Data Sets}

Up to now, we have concentrated on two specific choices of the {\em Planck}
data chains. Namely TTTEEE+lowE+lensing, for both $N_\nu$ fixed and variable.
In this section we compare our results with other chains made available by {\em Planck} as well as with the 2015 {\em Planck} data \cite{Planck2015}. 

\begin{center}
\begin{figure}[!htb]
\includegraphics[width=0.80\textwidth]{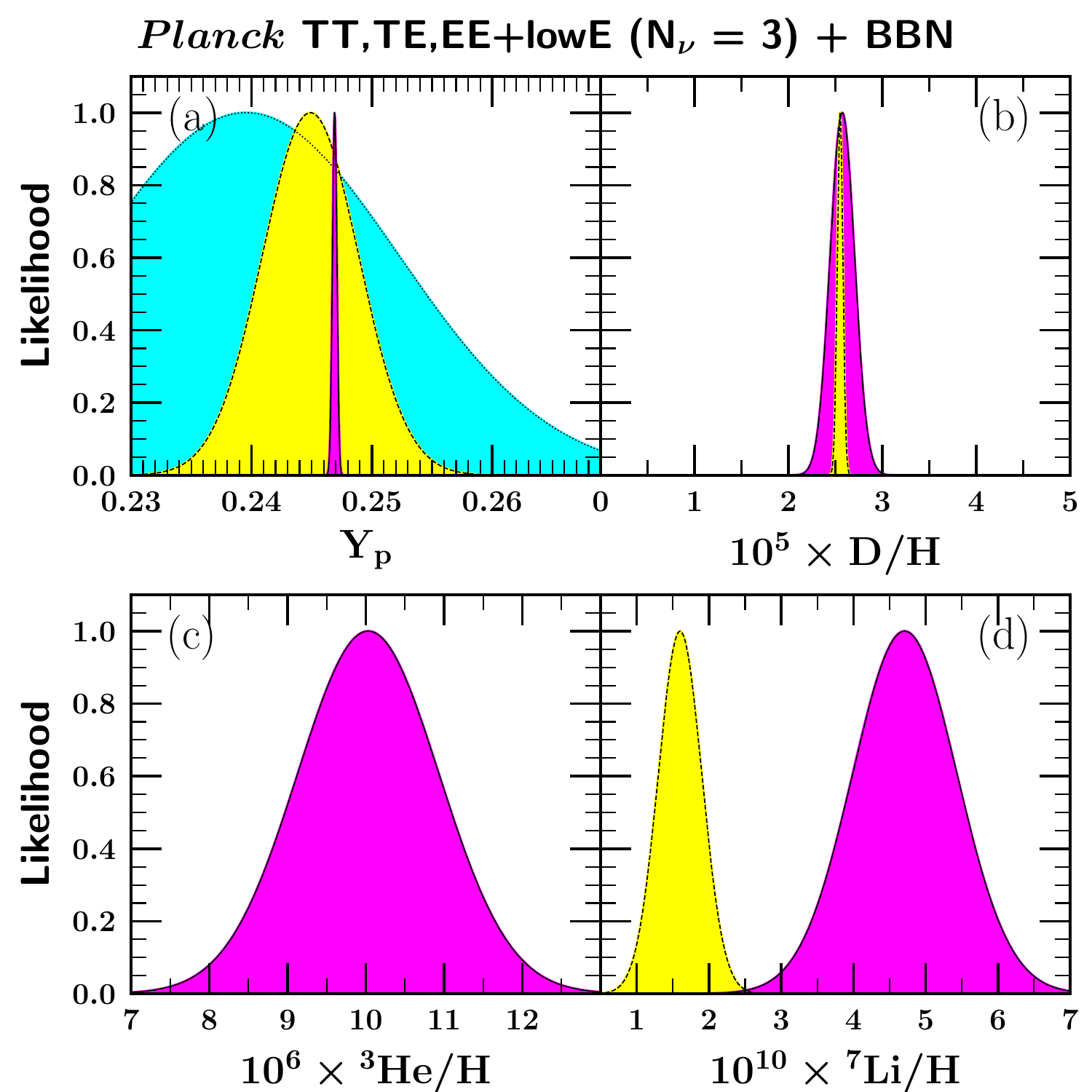}
\caption{ As in Fig.~\ref{fig:2x2abs_2d}, the
light element predictions using the CMB determination of the 
cosmic baryon density when CMB lensing is not included. }
\label{fig:2x2abs_2dnolens}
\end{figure}
\end{center}

We start by showing the likelihoods as  functions of each of the light elements
when lensing is not included in the CMB likelihood. These results, shown in Fig.~\ref{fig:2x2abs_2dnolens}, are very similar to those shown in Fig.~\ref{fig:2x2abs_2d} when lensing is included.  We see only a slight shift in the peak of the CMB likelihood as a function of $Y_p$. We have also determined the likelihood functions when
BAO data is added to the CMB chains, but again other than a slight shift in the CMB likelihood for \he4, the results are very similar and not shown here.

Given these results with no lensing, we
can cross correlate the likelihoods for each of the light elements.  We do this using the CMB MCMC chains directly as input to the BBN calculation.   For our purposes the tractable cases are where the points in each chain have integer weights.  This is the case for the analyses involving only {\em Planck} data and that do not combine other cosmological data such as lensing.  Fortunately, we have seen that the inclusion of lensing has a tiny effect on our results.  

Our procedure is to use the chains for the TTTEEE+lowE case (i.e., baseline except for no lensing) with fixed $N_\nu=3$ ($\neff = 3.045$).  Each point $k$ in the chain provides a CMB $(\eta_k,Y_k)$ and a weight ($w_k$). We run the BBN code $w$ times per point, sampling the nuclear reaction uncertainties randomly as usual.  The resulting set of abundances $X_k(\eta_k)$ at each point (properly weighted) follows the CMB+BBN likelihood in Eq.~(\ref{CMB-BBN}).  Moreover, we can examine the results for correlations between the light element predictions including the full range of nuclear and cosmological dependences.

\begin{figure}[!htb]
    \centering
    \includegraphics[width=0.7\textwidth]{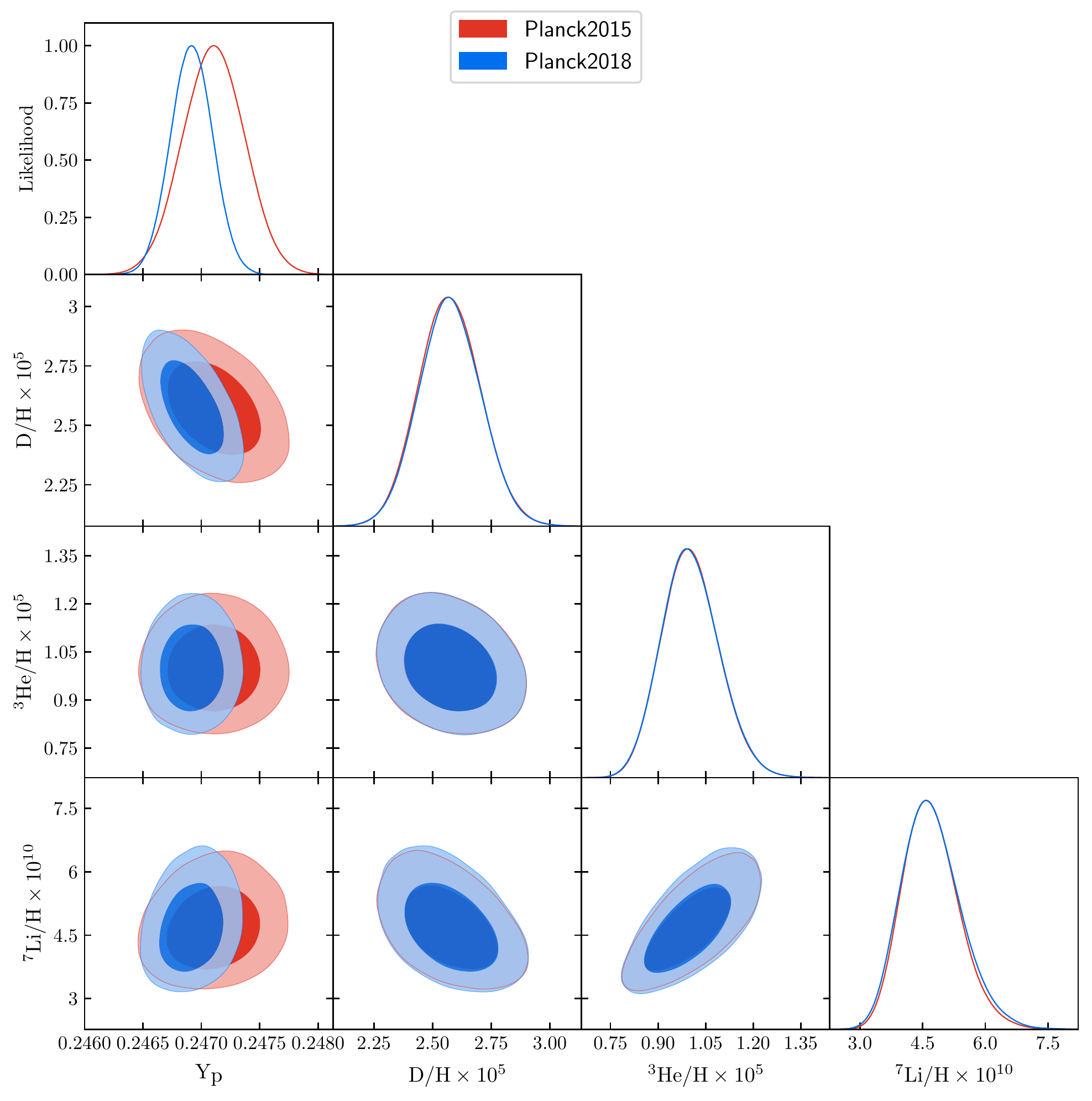}
    \caption{Light element likelihoods using the {\em Planck} MCMC chains for the TTTEEE+lowE case with $N_\nu=3$ and no lensing data. Off-diagonal plots show the correlations between light-element abundance predictions, for example the anticorrelation between D/H and \li7/H predictions.  Diagonal plots show the likelihood for each element, as in Fig.~\ref{fig:2x2abs_2dnolens} and akin to Fig.~\ref{fig:2x2abs_2d} but without lensing data in the CMB analysis.  Dark (light) blue regions:  68\% (95\%) CL for {\em Planck} 2018 data.  Underlying dark (light) red regions:   same but for {\em Planck} 2015 data.}
    \label{fig:triangle-TTTEEEnolensing}
\end{figure}

Our results for light element cross-correlations appear in Fig.~\ref{fig:triangle-TTTEEEnolensing}.
In most cases, there is in fact little correlation.
Some negative correlation between $Y_p$
and D/H is seen, as well as between
D and \li7. There is more evident positive correlation between \he3 and \li7.  These trends have been discussed in \cite{fiorentini}, and trace back to the underlying abundance dependence on the nuclear rates as seen in Eqs.~(\ref{yfit})-(\ref{li7fit}) and Table \ref{tab:sens}.  For example, the D--\li7 anti-correlation arises largely due to their common dependence on $d(p,\gamma)\he3$ and $d(d,n)\he3$, which destroy  D but produce \he3 that leads to \li7 through $\he3(\alpha,\gamma)\be7$.

Also seen in Fig.~\ref{fig:triangle-TTTEEEnolensing}, is a comparison between the current {\em Planck} 2018 results \cite{Planck2018}
with previous {\em Planck} 2015 results \cite{Planck2015}. Only in the case of \he4 is there any noticeable change in the likelihood function.

\begin{figure}[!htb]
    \centering
    \includegraphics[width=0.7\textwidth]{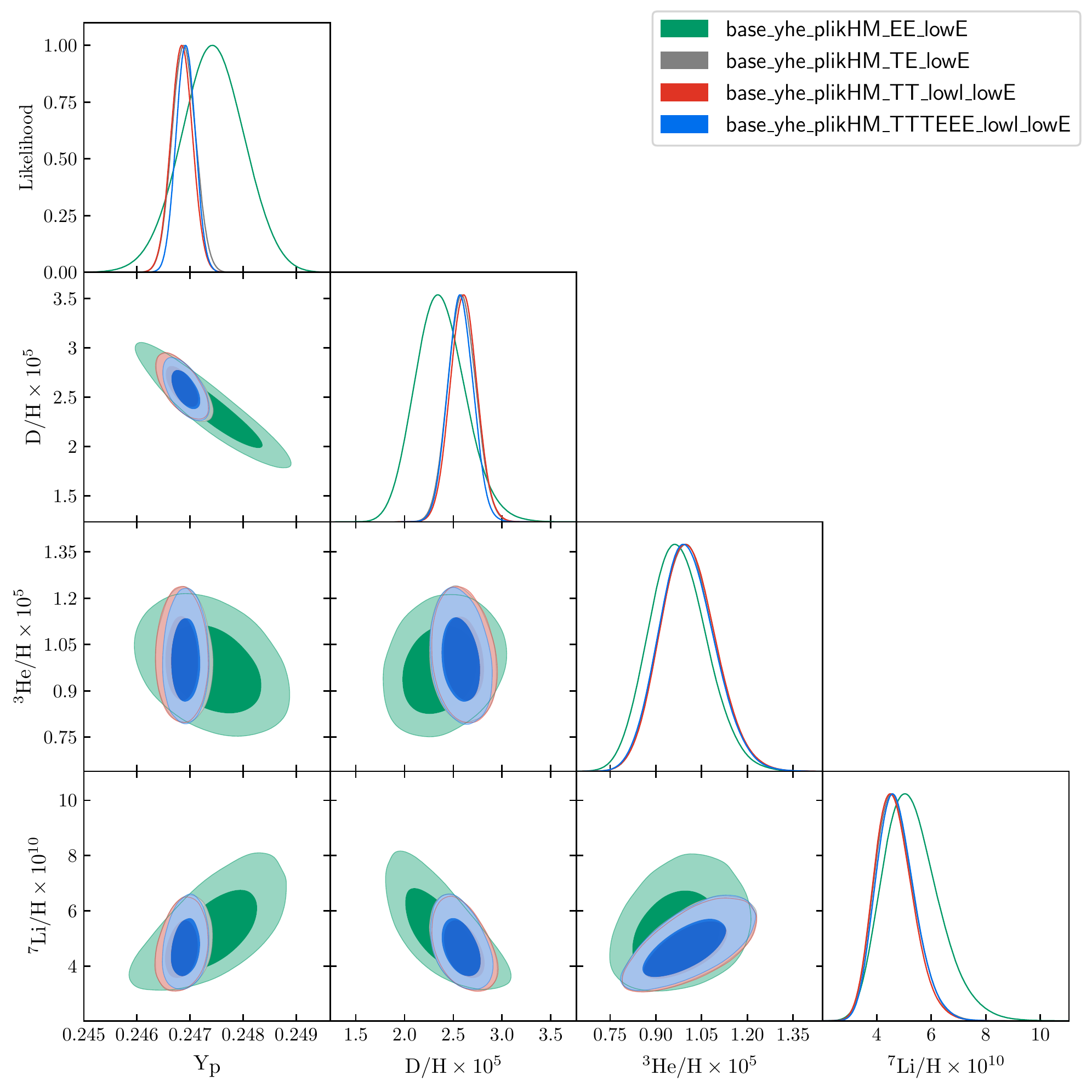}
    \caption{As in Fig.~\ref{fig:triangle-TTTEEEnolensing}
    comparing results using different {\em Planck} chains. Included here is the TTTEEE chain already shown in  Fig.~\ref{fig:triangle-TTTEEEnolensing}
    as well as the TT, TE, and EE chains. All include lowE data. }
    \label{fig:triangle-TTTEEEcompare}
\end{figure}

Another comparison between the {\em Planck} 
data chains is seen in Fig.~\ref{fig:triangle-TTTEEEcompare}.
Here we compare the 2018 likelihoods seen in  Fig.~\ref{fig:triangle-TTTEEEnolensing} for TTTEEE+lowE data, with the likelihoods based on TT+lowE, TE+lowE, and EE+lowE data separately. We first note that
likelihood functions for each of the light elements are extremely robust so long as some CMB temperature information is included. In fact, it is rather amazing that even when all temperature data is discarded using only the EE polarization data, we find reasonably accurate likelihood functions.

A comparison of the mean value predictions for each of the light elements depending on the particular {\em Planck} data set chosen is summarized in Table \ref{tab:cosm}. These correspond to the distributions shown along the diagonal in Fig.~\ref{fig:triangle-TTTEEEcompare} (for data columns 1-4) and in Fig.~\ref{fig:2x2abs_2d} (data column 5).
We have not shown the distributions
from column 6, but as one can see, it does not differ appreciably from column 5. 

\begin{table}[!htb]
\footnotesize
\caption{CMB+BBN predictions for the light element abundances when different sets of {\em Planck} data are used. All choices include lowE data. 
\label{tab:cosm}
}
\vskip .1in
\begin{tabular}{|l|c|c|c|c|c|c|}
\hline
 CMB+BBN & TT & TE & EE & TT+TE+EE &
 ...+lensing & ...+lensing +BAO\\
\hline
$Y_p$ & $0.2468 \pm 0.0002$ & $0.2469 \pm 0.0002$ & $0.2474 \pm 0.0006$ & $0.2469 \pm 0.0002$ & $0.2469 \pm 0.0002$ & $0.2470 \pm 0.0002$\\
\hline
$10^5$ D/H & $2.61 \pm 0.14$  & $2.59 \pm 0.14$ &  $2.37\pm0.26$ & $2.58 \pm 0.13$ & $2.57 \pm 0.13$ & $2.56 \pm 0.13$ \\
\hline
$10^5$ \he3/H & $1.01\pm 0.09$ & $1.01\pm 0.09$ & $0.97 \pm 0.09$ & $1.00\pm 0.09$ & $1.00\pm 0.09$ & $1.00\pm 0.09$  \\
\hline
$10^{10}$ \li7/H & $4.65 \pm 0.71$ & $4.69 \pm 0.72$ &  $5.30 \pm 1.03$ & $4.72 \pm 0.71$ & $4.72 \pm 0.72$  & $4.76 \pm 0.73$  \\
\hline
\end{tabular}
\end{table}

\section{The Future Impact of CMB-Stage 4 Measurements}
\label{S4}

CMB observations are poised to improve dramatically beyond their already impressive precision.
The next generation of ground-based CMB measurements is known as Stage 4 (CMB-S4),
with planning well underway \cite{CMB-S4,CMB-S4-RefDes}.  Being ground-based, these will dramatically improve the CMB precision at small angular scales and high multipoles.  Thus 
we can expect only incremental refinements in the precision of $\omb$ and thus $\eta$,
which is determined by the undamped acoustic peaks at large angular scales.  
For our forecasts, we will not assume any improvement over the {\em Planck} sensitivity 
to cosmic baryons.

The dramatic effect of CMB-S4 will be the improvements in $N_\nu$ and $Y_p$.  As noted in \S \ref{sect:cmbpars}, these parameters come from the CMB damping scale which is best accessed
through ground-based measurements.  Moreover, the CMB determination of $\neff$ is a 
science driver for CMB-S4.  In the standard cosmology, both the baryon-to-photon ratio and helium abundance should not change between the BBN and CMB epochs, so that it is meaningful to use the BBN $Y_p(\eta_{\rm cmb})$ relation.  When this is done, the constraints on $\neff$ are much stronger, and the CMB-S4 target sensitivity is $\sigma_{\rm S4}(\neff|{\rm BBN})=0.030$!  If this can be achieved,
then it would be possible to resolve the difference $\neff - N_\nu = 0.045$ due to neutrino heating.
This would offer a new probe of neutrino interactions during the BBN epoch.

The CMB-S4 sensitivity will depend on the total
sky coverage. To estimate the impact of CMB-S4, and for a survey covering a fraction $f_{\rm sky} = 0.5$ of the sky, we can infer sensitivity 
to $Y_p$ and $N_\nu$ from ref.~\cite{CMB-S4}.
We start with the {\em Planck} likelihood chain which gives ${\mathcal L}_{\rm CMB}(\eta,Y_p, N_\nu)$. For each point in the chain, we reduce the 
spread in $Y_p$ by 0.005/0.013
and in $N_\nu$ by 0.09/0.3. Then the \he4 and neutrino counting should have precisions of about 
\beq
\label{eq:S4-errors}
\sigma_{\rm S4}(\neff) \simeq 0.09 \qquad \qquad 
\sigma_{\rm S4}(Y_p) \simeq 0.005
\eeq
which matches the CMB-S4 forecasts.
We adopt these estimates, and now explore the impact on
our CMB-BBN joint analysis.

We start by once again showing the likelihoods
as functions of each of the light elements where for now, we assume that $N_\nu = 3$ is fixed. The main impact in this case is, unsurprisingly,  on \he4. 
The CMB-BBN likelihood ${\mathcal L}_{\rm CMB-BBN}(Y_p)$ shaded purple is identical as that shown previously (as the mean CMB prediction for $\eta$ has not changed and the uncertainties in $Y_p$ do not enter in this integration). Similarly, the 
observational likelihood (shaded yellow) and the {\em Planck} 2018 likelihood for \he4 is as before.  However, here we also show shaded in blue the expected improvement over the {\em Planck}
likelihood function for \he4 with CMB-S4.
Because we kept the mean of the distribution  with respect to $\eta$ fixed, we see only a reduction in the width of the distribution
which is now comparable with the observational uncertainty.  This is in fact a remarkable statement. The primordial \he4 abundance 
is expected to be equally well measured from
CMB experiments as it is from optical (and IR)
measurements made on extragalactic HII regions.

\begin{figure}[!htb]
    \centering
    \includegraphics[width=0.8\textwidth]{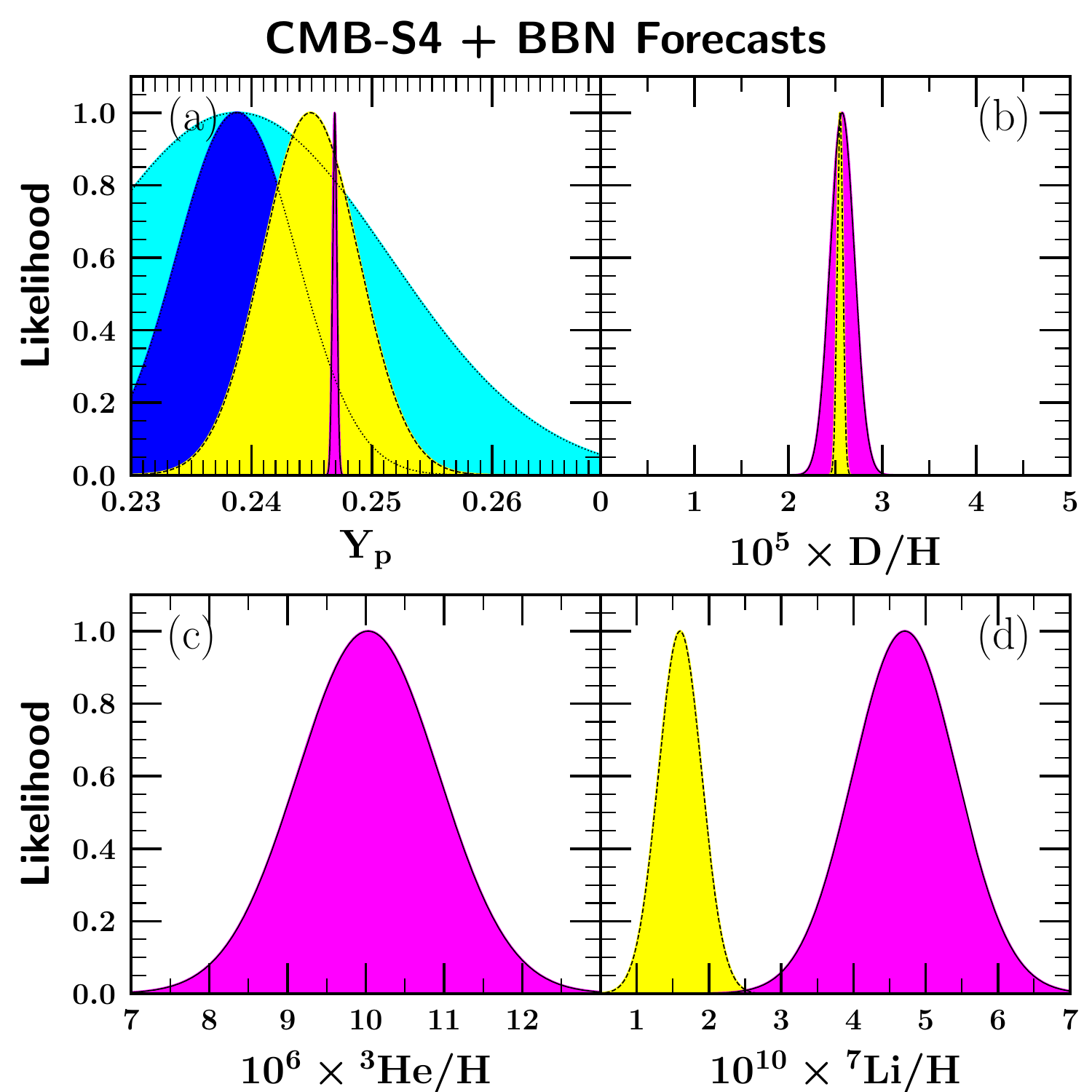}
    \caption{CMB-S4 forecasts for light element abundances with $N_\nu$ fixed, and with the CMB-S4 determination of \he4 (dark blue) using $\sigma_{\rm S4}(Y_p)=0.005$.  We see this is competitive with the astrophysical \he4 measurements. }
    \label{fig:CMB-S4-2x2like}
\end{figure}

\begin{figure}[!htb]
    \centering
    \includegraphics[width=0.9\textwidth]{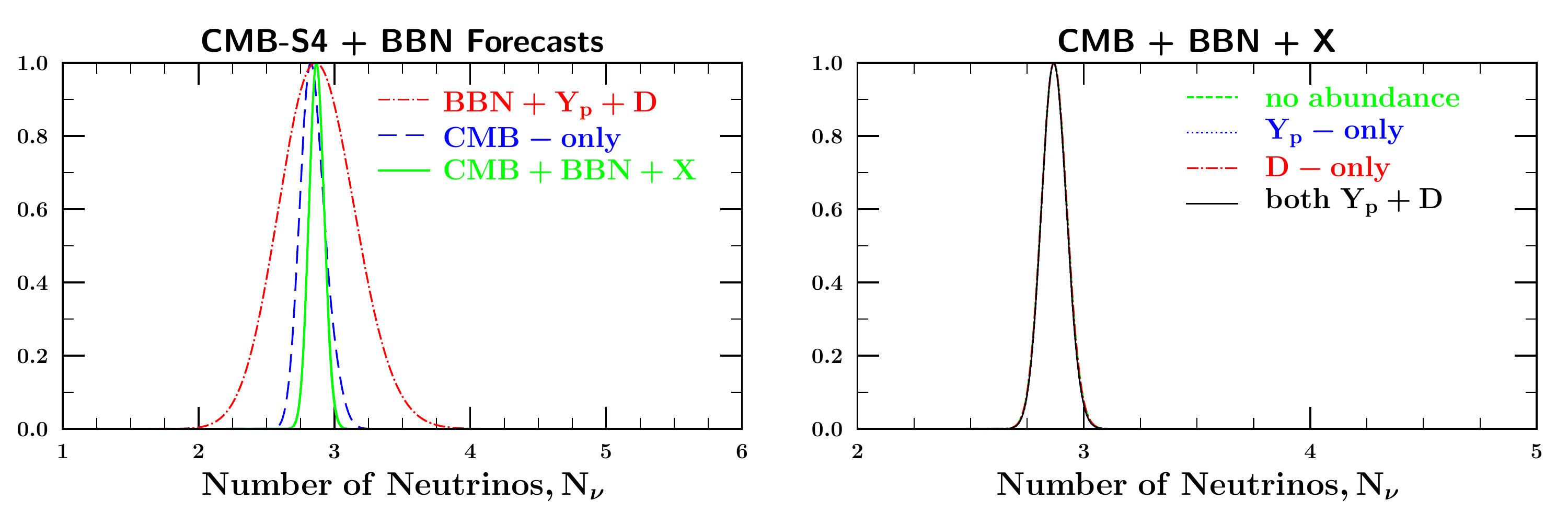}
    \caption{As in Fig.~\ref{fig:CMB-S4-nnu}, the CMB-S4 forecasts for $N_\nu$.}
    \label{fig:CMB-S4-nnu}
\end{figure}

The likelihood distributions shown in Fig.~\ref{fig:NBBN-nnu-baseline} as a function of $N_\nu$
are repeated in Fig.~\ref{fig:CMB-S4-nnu} assuming the CMB-S4 improvement in 
accuracy. The BBN+$Y_p$+D likelihood is of course
unchanged. But all of the distributions
based on CMB input have widths which are reduced by roughly a factor of 3. The mean values of $\eta$ 
and $N_\nu$ as well as the uncertainties in the means are summarized in Table \ref{tab:etannuS4}. Thus without a shift in the mean value of $N_\nu$, the 95\% confidence limit on $N_\nu -3$ becomes 0.036 from CMB-S4 alone, and the combined 95\% limit
becomes $N_\nu < 2.979$.  We note that SPT-3G expects to be able to place an upper limit on $N_\nu$ of 3.15 \cite{SPT-3G}.

\begin{table}[!htb]
\caption{The expected sensitivities to the baryon-to-photon ratio and effective number of neutrinos from CMB-S4, assuming unchanged mean values. 
\label{tab:etannuS4}
}
\vskip .1in
\begin{tabular}{|l|c|c|}
\hline
 Constraints Used & mean $\eta_{10}$ &  mean $N_\nu$  \\
\hline
CMB-S4 only & $6.090\pm0.060$ &  $2.846 \pm 0.095$  \\
\hline
\hline
CMB-S4+BBN & $6.086\pm 0.061$ &  $2.870 \pm 0.057$  \\
\hline
CMB+BBN+$Y_p$ & $6.086 \pm 0.060$ &   $2.869 \pm 0.055$  \\
\hline
CMB+BBN+D & $6.086 \pm 0.060$ &  $2.870 \pm 0.057$  \\
\hline
\hline
CMB+BBN+$Y_p$+D & $6.085 \pm 0.059$ &  $2.869 \pm 0.055$  \\
\hline
\end{tabular}
\end{table}

\begin{figure}
    \centering
    \includegraphics[width=0.7\textwidth]{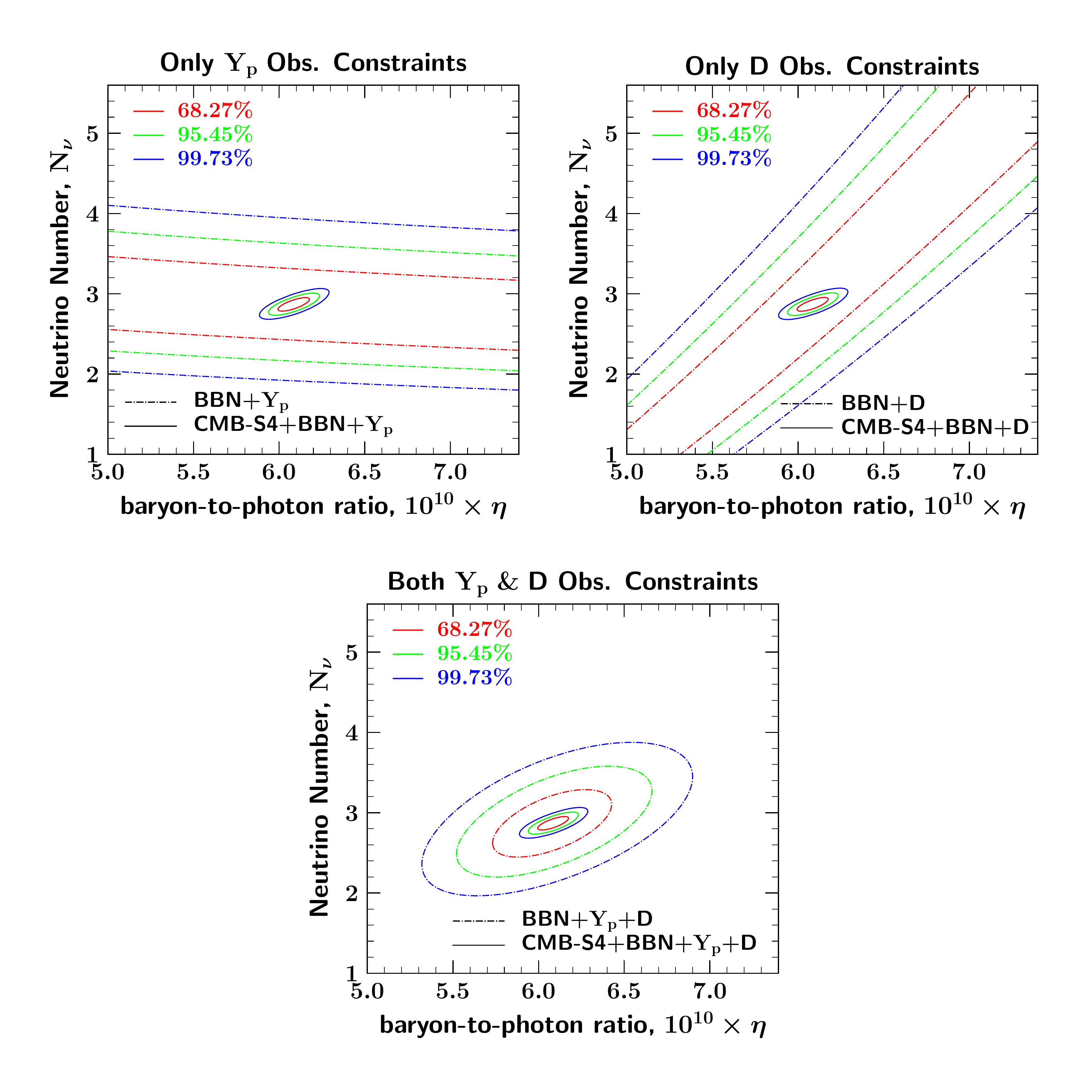}
    \caption{CMB S4 joint predictions for $\eta$ and $N_\nu$. }
    \label{fig:CMB-S4-eta-Nnu}
\end{figure}

Finally, the two-dimensional likelihood functions shown in Fig.~\ref{fig:2deta} are shown in Fig.~\ref{fig:CMB-S4-eta-Nnu} assuming the CMB-S4 improvements given in Eq.~(\ref{eq:S4-errors}). 
The BBN plus observations likelihoods are of course unchanged, but we see a dramatic improvement when 
the CMB likelihood is included. The CMB+BBN+OBS closed
loops are now significantly smaller in each of the three panels. Thus there is good reason to eagerly await these results.

\section{Discussion}
\label{sec:disc}

There is no question that starting with WMAP \cite{wmap},
BBN results and predictions changed both quantitatively as well as qualitatively. Rather than loosely fitting the baryon-to-photon ratio from observations of the light element abundances, WMAP first obtained this ratio to high accuracy,
providing a very stringent test of BBN and well as physics
beyond the Standard Model. {\em Planck} 2015 \cite{Planck2015}
and then later {\em Planck} 2018 \cite{Planck2018} further increased
the accuracy of BBN inputs. We fully expect that CMB-S4 \cite{CMB-S4} will continue this trend of
improved accuracy and precision.

It is important to note as we have argued throughout this work,
some of the accuracy attained by CMB measurements is reliant on BBN.  For example, the \he4 abundance, $Y_p$ is often used in the {\em Planck} likelihood functions. As we have seen, in the absence of any BBN relation between $\omb$ and $Y_p$,
we are able to obtain  1\% accuracy in $\eta$.  When the BBN relation is used (here we have used our own BBN generated relation in the {\em Planck} likelihood distributions), the accuracy improves to less than 0.6\%. This can be compared with the roughly 3\% accuracy attainable through BBN and D/H abundance
determinations. 

Similarly, using {\em Planck} 2018, the CMB alone can only determine
the number of light degrees of freedom (where we use $N_\nu$
as a surrogate) to an accuracy of about 11\%.
This is only marginally worse than BBN can do on its own
using abundance data, which fixes $N_\nu$ to just under 10\%. 
Adding BBN (but no abundance data) to the CMB likelihood chain, improves this accuracy to less than 7\% and using abundance data
(dominated in importance by \he4) drops the uncertainty in $N_\nu$ to 5.3\%. 

Perhaps most impressive when we look forward is the projected 
accuracy to which the \he4 abundance will be determined by CMB-S4.  We expect this will compete with direct measurements
of \he4 emission lines from extragalactic HII regions,
thus providing a new and key test of BBN and Standard Model physics. 

On the side of nuclear physics, new data is needed to resolve the theory/experiment discrepancy in the $d(p,\gamma)\he3$ rate.
Here, we have (in our baseline results) relied on existing
nuclear data using measurements in the energy range of interest to BBN. Instead, using data at lower energy (where more accurate data is available), and using theory to extrapolate to BBN energies, leads to an increased reaction rate, and therefore a slight paucity in D/H produced in the early Universe as compared with abundance determinations from high redshift quasar absorption systems. A direct measurement of the absolute 
$d(p,\gamma)\he3$ cross section at energies most relevant to BBN would be of great value. 
Recent steps in this direction have been made \cite{tisma},
though the interpretation of these results are not yet conclusive. Indirect approaches such as the Trojan Horse method could also be useful since they are suited to probe the shape but not the absolute normalization of the cross section.  Here this could be useful in smoothly connecting measurements at
energies above and below the BBN energy window, but the challenge would be to achieve the needed precision.

As much as the concordance between BBN theory, light element abundance observations, and measurements of the CMB is impressive, the elephant in the room is the \li7 abundance.
There is a clear mismatch between the BBN predicted abundance
and the abundance seen in metal-poor stars \cite{cfo5}. We have largely ignored this discrepancy as its not clear whether its resolutions lies in nuclear physics, beyond the Standard Model
particle physics, the observational determination, or the
stellar history of \li7. Obviously, a convincing resolution of this vexed problem would strengthen the power of BBN as the earliest probe of the Universe based on Standard Model physics.

\appendix

\section*{Appendix}
\label{app}

\section{Useful Conversions}
\label{sect:convert}

Different measures are used to quantify the cosmic baryon content and helium abundances.
For convenience we summarize them here, in a compact formalism.

\subsection{Helium Abundances:  Baryon and Mass Fractions}

Helium-4 abundances--both observed and predicted--are determined in terms of number densities,
but traditionally are reported in two
closely related ways, defined as follows.  Let nuclide $i$ with mass number (and baryon number) $A_i$, have number density $n_i$.  Its baryon number density is thus $n_{{\rm B},i} = A_i n_i$, and so
the total baryon density is $n_{\rm B} = \sum_j A_j n_j$.  Due to baryon conservation,
$n_{\rm B}$ is left unchanged in nuclear reactions, and thus the comoving baryon density $n_{\rm B} a^3$ is constant.

The {\em baryon or nucleon fraction} (or in a bit of a misnomer, the {\em baryonic mass fraction}) in species $i$ is naturally defined to be
\beq
\bfrac_{{\rm B},i} = \frac{A_i n_i}{\sum_j A_j n_j} \, ,
\eeq
which is invariant under cosmic expansion.
If we only consider the most abundant species \iso{H}{1} and \he4, then we use the 
traditional shorthand
\beqar
X \equiv X_{\rm B,1} & = & \frac{n_1}{n_1 + 4n_4} \\
Y \equiv X_{\rm B,4} & = & \frac{4 n_4}{n_1 + 4 n_4} \, ,
\eeqar
where clearly $X+Y=1$.   

A related but distinct abundance measure comes from considering the
{\em mass densities} in our set of nuclides.  Let species $i$ have mass $m_i$.
Then its mass density is $\rho_i = m_i n_i$, and the total baryonic mass
density is $\rho_{\rm B} = \sum_j \rho_j$.  This naturally leads us to
define a {\em true mass fraction} 
\beq
\mfrac_{{\rm M},i} = \frac{\rho_i}{\rho_{\rm B}} = \frac{m_i n_i}{\sum m_j n_j} \, .
\eeq
Of course, to a first approximation, all nucleons have roughly the same ``baryonic mass'' $m_{\rm B}$,
in which case $m_i \approx A_i m_{\rm B}$, and then we find the the true mass
fractions and baryonic mass fractions are identical:  $\mfrac_{{\rm M},i} \approx \bfrac_{{\rm B},i}$.
For this reason, and to be consistent with earlier work, {\em outside of this section} we refer to the baryonic mass fraction as simply the mass fraction.

However, due to binding energies and the neutron-proton distinction, not all baryons have
the same mass.  It will prove convenient to choose hydrogen atoms (i.e., \iso{H}{1}) as the mass standard, 
since before, during, and after BBN, protons are the dominant species by number and mass,
and hence the mean baryonic mass is nearly the hydrogen mass.  
Thus for species $i$ we write
\beq
m_i \equiv A_i m_{\rm H} + Q_i \equiv A_i m_{\rm H} \left( 1 + \beta_i \right) \, ,
\eeq
which defines a ``hydrogenic mass excess'' $Q_i$, and a fractional excess
\beq
\beta_i = \frac{Q_i}{A_i m_{\rm H}} = \frac{\Delta_i - A_i \Delta_1}{A_i m_{\rm H}} \, ,
\eeq
where the final expression uses the usual mass excess or mass defect $\Delta_i$
defined 
$m_i = A_i m_u + \Delta_i$ using the amu $m_u$ as the mass standard.  
Note here that we use {\em atomic} masses, so there is an electron contribution $Z_i m_e$
for each atomic mass, and $m_{\rm H} = m_p + m_e$.

Notice that by definition, our hydrogen-based mass excess has $Q_1 = 0 = \beta_1$,
while $\he4$ has $Q_4 = \Delta_4 - 4 \Delta_1 < 0$, a negative value due to 
the \he4 binding. Thus the dimensionless binding correction is also negative: $\beta_4 = - 7.119 \times 10^{-3}$.

Using hydrogen as the mass standard makes our formulae particularly simple when we ignore all but 
\iso{H}{1} and \he4, adopting a shorthand
\beqar
\mfracX \equiv \mfrac_{{\rm M},1} & = & \frac{m_p n_1}{m_p n_1 + m_4 n_4} = \frac{n_1}{n_1 + 4 (1+\beta_4) n_4} \\
\label{eq:y}
\mfracY \equiv \mfrac_{{\rm M},4} & = & \frac{4 (1+\beta_4) n_4}{n_1 + 4 (1+\beta_4) n_4}  
\eeqar
We can use the last expression to write the helium mass fraction in terms of its baryon fraction
and vice versa:
\beqar
\label{eq:b2m}
\mfracY & = & \frac{(1+\beta_4) Y}{1 + \beta_4 Y} \\
\label{eq:m2b}
Y & = & \frac{\mfracY}{1+\beta_4(1-\mfracY)}
\eeqar
One can show that since $\beta_4<0$, then $Y > \mfracY$ always:  physically, the \he4 mass fraction
is lower than its baryon fraction due to its binding energy.
Numerically the difference is relatively small.  For example, a baryon fraction of $Y = 0.2450$
corresponds to a mass fraction $\mfracY = 0.2437$;
we look forward to the day when precision is sufficient that this distinction is important!

{\em Planck} reports \he4 abundances both as baryon fractions ($Y_p^{\rm BBN}$ in their notation)
and mass fraction (their $Y_p$). Our \he4 results are quoted as baryon fractions, and this
is how observed results are reported in, e.g., the Aver et al.~work of ref.~\cite{aos} and thereafter.

Finally, one could instead measure baryonic/nucleonic mass densities
and mass fractions
strictly including nuclei only, with no electron contribution.
In this case our analysis is changed to using the
proton mass $m_p$ as a standard,
and then writing the mass of nucleus of species $i$ as 
\beq
m_i^\prime = A_i m_p + Q_i^\prime = (1 + \beta_i^\prime) A_i m_p
\eeq
with $\beta_i^\prime = Q_i^\prime/A_i m_p$.
Ignoring atomic binding energies, we have
$m_i^\prime = m_i - Z_i m_e$, with $Z_i$ the charge or atomic
number of nuclide $i$.
This in turn means that
\beq
Q_i^\prime = Q_i + (A_i- Z_i) m_e = Q_i + N_i m_e 
\eeq
that is the two corrections are different due to the 
lack of the electrons associated with the number $N_i=A_i-Z_i$ of neutrons
in each nucleus.
Finally then we have
\beq
\beta_i^\prime = \frac{Q_i^\prime}{A_i m_p}
= \frac{m_{\rm H}}{m_p}\beta_i + \left(1-\frac{Z_i}{A_i}\right) \frac{m_e}{m_p}
\eeq
which gives $\beta_4^\prime = -6.8501 \times 10^{-3}$.
Using this we can define a nucleon-only mass density
$\rho_{\rm B}^\prime = \sum_j m_j^\prime n_j$,
and associated mass fraction
$\mfrac_i^\prime = m_i^\prime n_i/\rho_{\rm B}^\prime$.
And for \he4 we have the same relations as in eqs.~(\ref{eq:b2m})
and (\ref{eq:m2b}), with $\mfracY \rightarrow \mfracY^\prime$
and $\beta_4 \rightarrow \beta_4^\prime$.

\subsection{Measures of the Baryon Density}

The CMB probes the comoving
baryonic mass density $\rho_{\rm B} a^3 = \rho_{\rm B,0}$.  This is expressed
in the units of the present critical density $\rho_{\rm crit,0} = 3H_0^2/8\pi G_N$
as the baryon density parameter $\Omega_{\rm B} = \rho_{\rm B,0}/\rho_{\rm crit,0}$
and the reduced baryon density parameter $\Ombh2 \equiv \omb$ which removes the uncertainty
in the Hubble parameter $H_0 = h H_{100}$, with $H_{100} = 100 \ \rm km \ s^{-1} \ Mpc^{-1}$.  

The present baryonic mass density is related to the present baryon number density $n_{\rm B,0}$
via the average mass $\avg{m_{\rm B}}$ per baryon:
\beq
\rho_{\rm B,0} = \avg{m_{\rm B}} \ n_{\rm B,0}
\eeq
Writing this in terms of $\eta = n_{\rm B}/n_\gamma$, and using the blackbody relation
$n_\gamma = 2 \zeta(3)/\pi^2 \ (kT/\hbar c)^3$, we can relate the baryonic measures
\beqar
\omb & = & \frac{\avg{m_{\rm B}} \ n_{\rm B,0}}{h^{-2} \rho_{\rm crit,0}} \\
& = & \frac{16  \zeta(3) \ G_N \ m_{\rm H} \ (kT_*/\hbar c)^3}{3 \pi H_{100}^2} 
\pfrac{\avg{m_{\rm B}}}{m_{\rm H}} \pfrac{T}{T_*}^3 \ \eta \\
& = & 3.6594 \times 10^{-3} \ \pfrac{\avg{m_{\rm B}}}{m_{\rm H}} \pfrac{T}{2.7255 \ \rm K}^3 \ \eta_{10}
\eeqar
where we choose a fiducial CMB temperature to be the present value $T_* = 2.7255 \ \rm K$.

We see that the conversion depends on the mean mass per baryon, and hence on the
composition, as has been pointed out
by many authors and discussed in detail by Gary Steigman \cite{gary}.  Using our formalism
from the previous section, and taking the excellent approximation of
ignoring all but \iso{H}{1} and \he4, we have
\beq
\label{eq:mbar}
\avg{m_{\rm B}} = \frac{\sum m_i n_i}{\sum A_j n_j} = \left(1 + \beta_4 Y \right) m_{\rm H} 
\equiv (1+\delta) m_{\rm H} 
\eeq
and we see that the correction factor $\delta = \beta_4 Y <0$ since $\beta_4 < 0$:
helium binding reduces the mass per baryon. For a fiducial value of $Y_* = 0.245$
we have $\delta_* = - 1.744 \times 10^{-3}$, and we can write 
\beqar
\label{eq:delta}
\delta & = & \delta_* + \beta_4 (Y- Y_*) \\
& = &  -\left[ 1.744  +  7.119  \left( Y - 0.245 \right) \right] \times 10^{-3} \\
\frac{1+\delta}{1+\delta_*} & = & 1 - 7.131 \times 10^{-3} \left(Y-0.245 \right)
\eeqar
Using this we arrive at the conversions
\beqar
\label{eq:ombat}
\omb & = & \frac{16  \zeta(3) \ G_N \ (1+\delta) m_{\rm H} \ (kT_*/\hbar c)^3}{3 \pi H_{100}^2} 
 \pfrac{T}{T_*}^3 \ \eta \\
 \label{eq:omb2eta_at}
  & = & 3.6529 \times 10^{-3} 
  \ \left[ 1 - 7.131 \times 10^{-3} \left(Y-0.245 \right) \right]
  \ \pfrac{T}{2.7255 \ \rm K}^3 \ \eta_{10} \\
  \label{eta2omb_at}
\eta_{10} & = &  273.754 
\ \left[ 1 - 7.131 \times 10^{-3} \left(Y-0.245 \right) \right]^{-1}
\ \pfrac{2.7255 \ \rm K}{T}^3 \ \omb
\eeqar
This agrees well with literature results, including ours in CFOY.

Note that if one measures \he4 by its true mass fraction, one simply adopts the conversion above (Eq.~\ref{eq:m2b}) to find the baryon fraction $Y$ for which these expressions are particularly simple.

Finally, as noted above we can consider the strictly baryonic
and thus nucleonic mass density, which excludes the
electron mass contribution.
Then we use as our mass standard the proton $m_p$,
and so the mean nuclear mass in eq.~(\ref{eq:mbar})
becomes $\avg{m_{\rm B}} = m_p (1+\beta_4^\prime Y)$,
which gives the correction term $\delta^\prime = \beta_4^\prime Y$
which leads to 
\beqar
\omb^\prime & = & \frac{16  \zeta(3) \ G_N \ (1+\delta^\prime) m_p \ (kT_*/\hbar c)^3}{3 \pi H_{100}^2} 
 \pfrac{T}{T_*}^3 \ \eta \\
  & = & 3.6512 \times 10^{-3} 
  \ \left[ 1 - 6.862 \times 10^{-3} \left(Y-0.245 \right) \right]
  \ \pfrac{T}{2.7255 \ \rm K}^3 \ \eta_{10} \\
\eta_{10} & = &  273.885
\ \left[ 1 - 6.862 \times 10^{-3} \left(Y-0.245 \right) \right]^{-1}
\ \pfrac{2.7255 \ \rm K}{T}^3 \ \omb^\prime
\eeqar
which are the analogs of eqs.~(\ref{eq:ombat})--(\ref{eta2omb_at}).

\section{Fits to the $\be7(n,p)\li7$ Rate}

\label{sect:7benp-fits}

The $\be7(n,p)\li7$ rate is one of the two dominant contributions to the BBN \li7 uncertainty budget.  Measurements of this reaction have taken two approaches.  One is to use the inverse reaction $\li7(p,n)\be7$ and invoke detailed balance to infer the forward rate.  This approach is much easier experimentally, as the beam and target are both stable.  The drawback is that the low-energy behavior of the forward reaction, which is of the most interest, requires careful measurement right at the threshold energy of the inverse reaction.  The alternative is to measure the forward reaction at the cost of producing a radioactive target and a neutron beam.   

Both the direct and inverse method has been pursued over the years.  Most recently, the n\_TOF experiment at CERN has added new data on the forward reaction \cite{damone2018}.  
We combined this forward-kinematics
determination of $\be7(n,p)\li7$ with existing and largely inverse kinematics data,
  appropriately weighting for statistical and systematic uncertainties, following the procedure in \cite{cyburt}.  The resulting rate $R(E) = N_{\rm A} \ \sigma(E) \ v$
  in energy space, is plotted in Fig.~\ref{fig:be7np}.  
  
  As noted in \S\ref{sect:7benp}, the n\_TOF data lie above
  the Sekharan data \cite{sekharan} near BBN energies,
  and much above the Koehler data~\cite{koehler} at low energies.
  This points to systematic errors which will translate
  into increased discrepancy errors in our fits.
  The extent of the discrepancy is energy-dependent
  and thus we will explore the uses of different energy ranges.

\begin{figure}[!htb]
    \centering
    \includegraphics[width=0.7\textwidth]{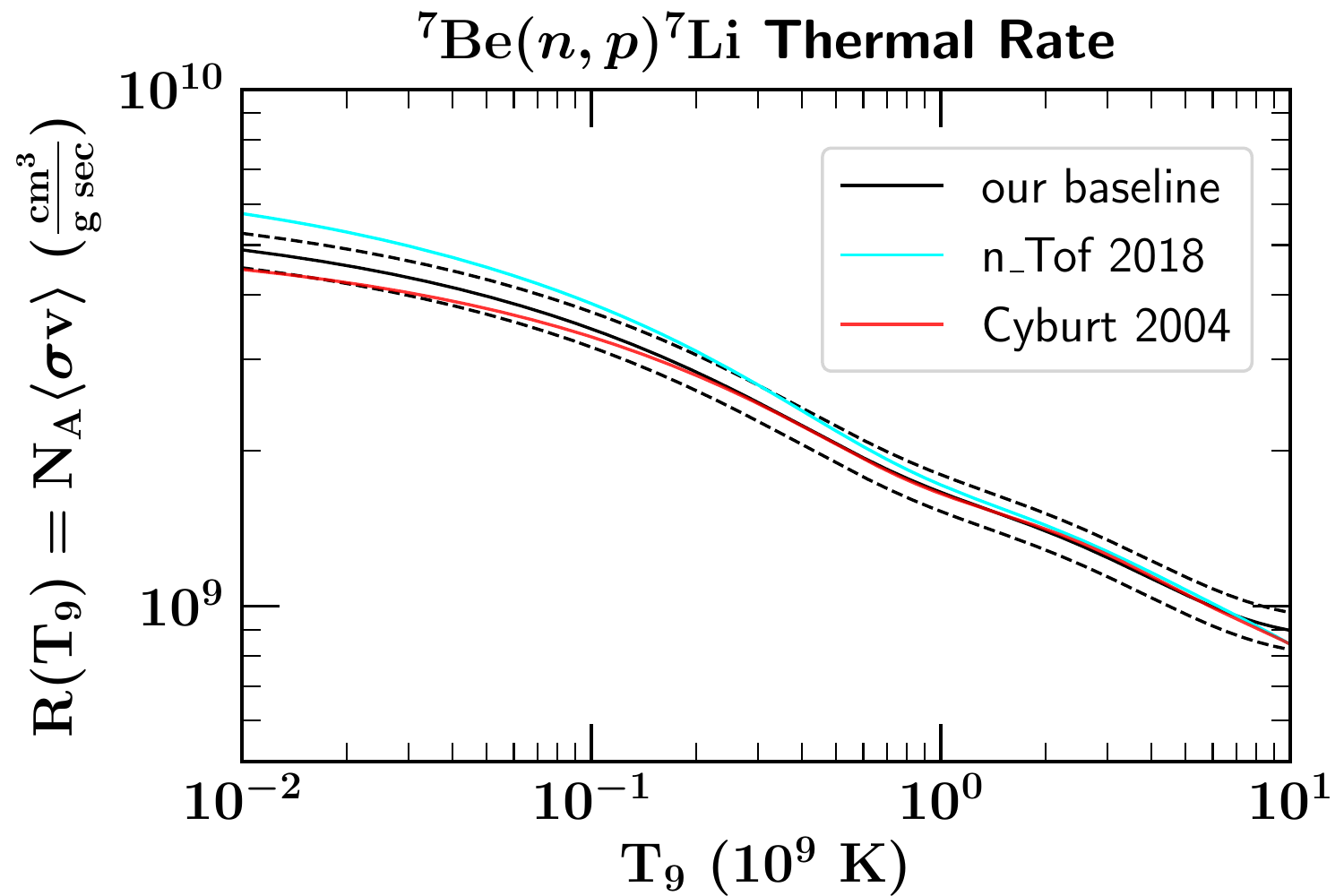}
    \caption{Thermal rate for $\be7(n,p)\li7$.
    Our rate is the thick black line, with the $1\sigma$ error range shown in dotted lines.  For comparison we also show in red the results for Cyburt \cite{cyburt}, and in cyan the n\_TOF recommended rate  \cite{damone2018}. }
    \label{fig:thermalrate}
\end{figure}

We numerically calculate the thermal average rate $R(T) = N_{\rm A} \ \avg{\sigma v}$, and plot the results in Fig.~\ref{fig:thermalrate}.  The gradual decline with temperature reflects the drop in the reaction rate $R(E)$ versus energy in Fig.~\ref{fig:be7np}.  

Fig.~\ref{fig:thermalrate} also shows the rates for Cyburt \cite{cyburt}, which lacked the n\_TOF data.  We also show the n\_TOF recommended rate tabulated in \cite{damone2018}, which uses their own data along with that of Sekharan \cite{sekharan} but not the other data in Fig.~\ref{fig:be7np}.  We see that the rates diverge at low temperature, reflecting the discrepant data at low energy:  the Cyburt fit uses the higher data, n\_TOF the lower, and ours is a compromise with the price of a higher systematic uncertainty.  However, at the temperature $T_9 \sim 1$ relevant for \be7 evolution, the three rates are almost the same.  Thus we already expect that the different rate fits will give similar final \li7 abundances, but the  uncertainties in our fit to be larger.

\begin{center}
\begin{figure}[!htb]
\includegraphics[width=0.66\textwidth]{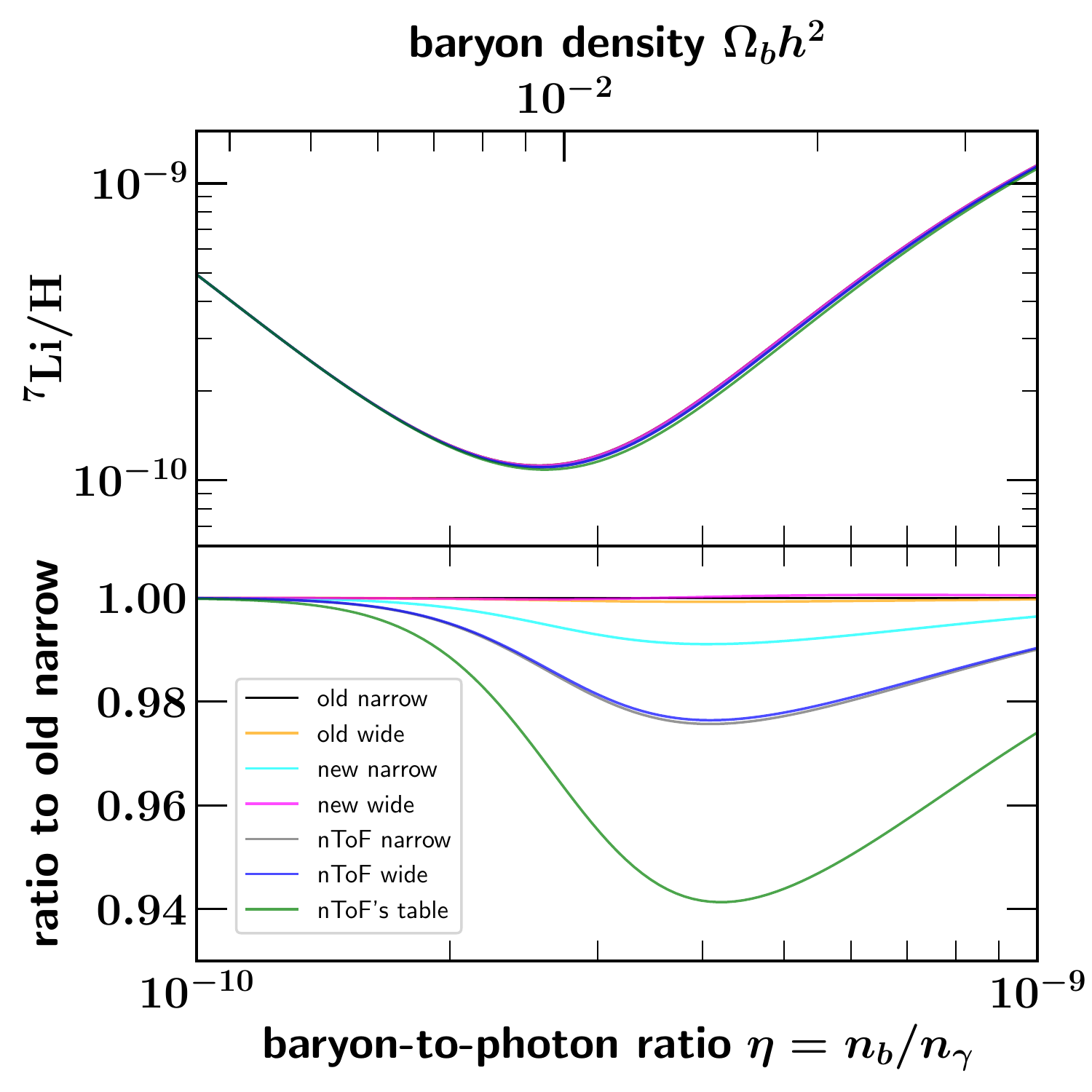} 
\caption{
The abundance of \li7/H as a function of
cosmic baryon content, comparing various input rates for \be7(n,p)\li7. In the lower panel, we the ratio of each rate considered relative to the old narrow rate used in CFOY. 
\label{fig:schramm77}
}
\end{figure}
\end{center}

Figure~\ref{fig:schramm77} shows the impact of  different $\be7(n,p)\li7$ rate choices on primordial \li7.  What is plotted are the central-value predictions for a set of different fits to the rates.  All of the curves use our fitting procedure, for which one must specify the data to be fit, and the energy range in the fit.  We consider both a ``narrow'' energy region $E_{\rm CM} \in (0.01,1.0) \ \rm MeV$ centered on the BBN, and a ``wide'' range that includes all data, particularly that at low energies.  These choices can give different results due to the influence of the lowest-energy points on our fit--these anchor the first two terms in our $R(E)$ polynomial.    
  
As expected, all of the choices give very similar results.  Indeed, the bottom panel of Fig.~\ref{fig:schramm77} shows the ratio of results using alternative rates relative to CFOY, which uses the Cyburt rate.   Most importantly, none of the predictions differ by more than about 6\%, which is slightly smaller than the systematic error.  And such differences are not significant for the lithium problem.

In detail, we see that all \li7 predictions lie below that from our old rate (old narrow), because our that has the least \be7 destruction.  Picking a larger energy range (old wide) does not change things, as there is no discrepancy at low energies.   We see that our new fiducial rate (new narrow) is slightly higher, by about 1\%, reflecting the slightly higher \be7 destruction. The largest change is for the n\_TOF tabulation which omits some of the data we use, and which has a higher low-energy and low-temperature rate.

\section{Impact of Other \be7 Reaction Rates}

\label{sect:other-rates-app}

\begin{figure}
    \centering
    \includegraphics[width=0.7\textwidth]{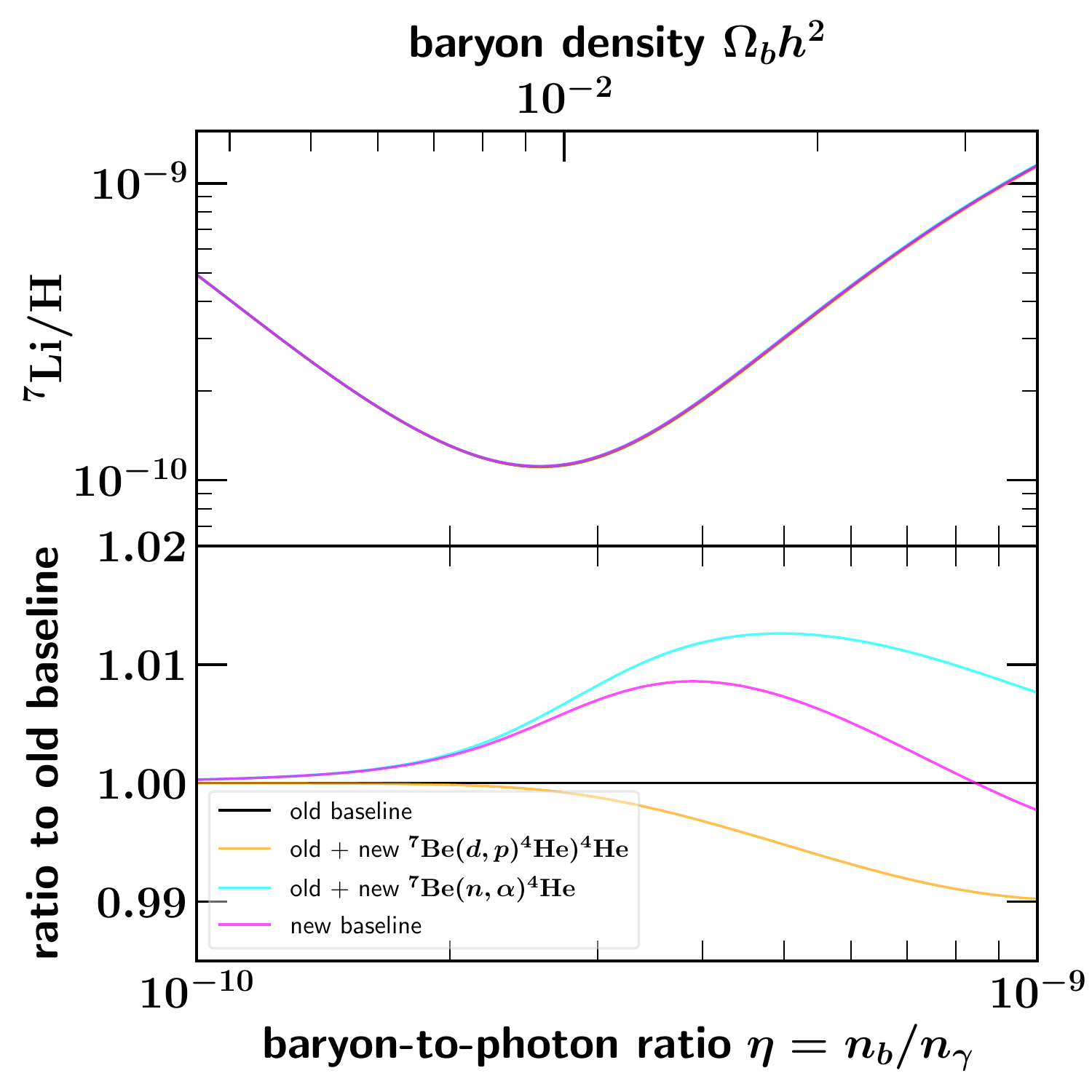}
    \caption{Lithium predictions vs $\eta$ as in bottom panel of Fig.~\ref{fig:schramm77}, showing the effect of updated rates for $\be7(d,p)\he4 \he4$ and $\be7(n,\alpha)\he4$.
    \label{fig:7be-other-rxns}
    }
\end{figure}

As discussed in \S \ref{sect:other-rates}, 
there have been recent measurements 
on rates contributing to \be7 destruction.  
A number of groups have firmed up the experimental 
picture of $\be7(n,\alpha)\he4$, finding that this
rate is smaller than the old estimates,
which implies that this already subdominant rate
has further diminished in importance.
On the other hand, recent $\be7(d,p)\li7$ data
have claimed that the rate has increased
to the point of having a non-negligible 
but still not dominant effect on \be7.  

We show the impact of the new rates
in Fig.~\ref{fig:7be-other-rxns}.
The top panel shows the impact of the new rates on the \li7 portion of the Schramm plot.
The changes are so small they are mostly within
the width of the curve.  Thus we have added
the bottom panel, showing the ratio of the new
to old \li7 
for the adopted rate updates, both separately and
together.  

We see that the new $\be7(n,\alpha)\he4$ rate leads to a \li7 increase of around 1\% in the $\eta$ range of interest.  This change is as expected, and also
agrees with expectations based on the rate scalings
from Eq.~(\ref{eqn:sens}) and Table \ref{tab:sens}.

In contrast, the increase in the $\be7(d,p)\li7$ rate lower \li7, albeit by a small amount--somewhat below 1\% at the CMB baryon density.  This is in agreement with our expectations, and those of \cite{gai}, but in contrast
to the claims of ref.~\cite{Rijal2018}.

The net effect is a slight ($\la 1\%$) change in \li7.   Obviously this small shift has no perceptible impact on the lithium problem.

\section{Dependence on D(p,$\gamma$)\he3}

\label{sect:dpg-app}

\begin{figure}
    \centering
    \includegraphics[width=0.7\textwidth]{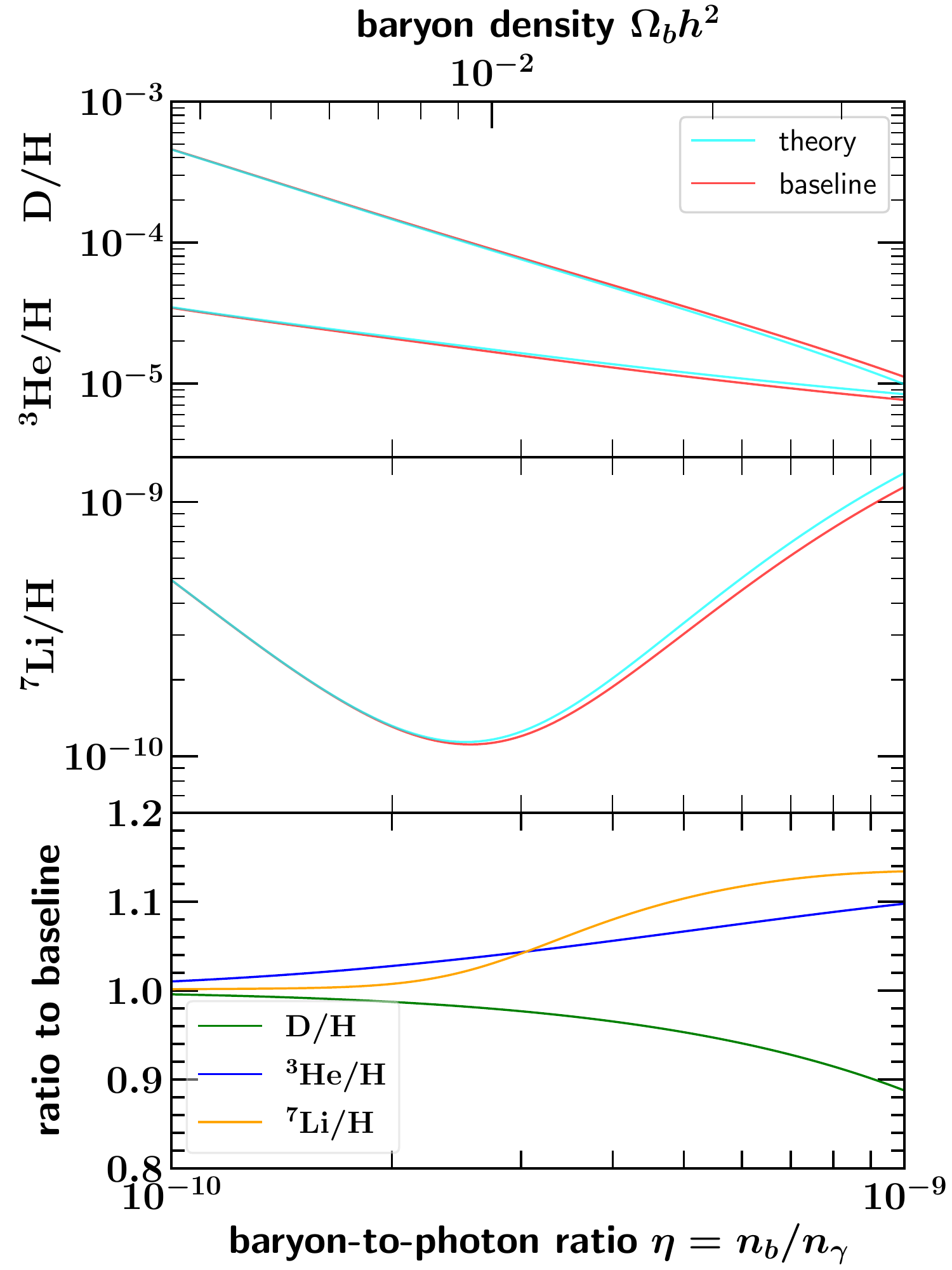}
    \caption{The effect of alternative theory-based approach to $d(p,\gamma)\he3$.  We adopt the rates of
    \cite{marc}.  Note that \li7/H shift makes the lithium
    problem slightly worse.  More importantly, this increases D/H
    enough to spoil the excellent agreement with observations.
    \label{fig:dpg}
    }
\end{figure}

As discussed in \S \ref{sect:dpg}, the $d(p,\gamma)\he3$ reaction
has been the source of some controversy of late.  
By construction, our adopted rates are empirical and thus
follow the experimental data.  
Theory-based fits give a higher rate at BBN energies and temperatures.

We show the effect of adopting the theory-based rates
in Fig.~\ref{fig:dpg}.   Here we show the change in abundances 
as a function of cosmic baryon content.  The shift in $Y_p$ is small
and are omitted from the plot.  The changes to D/H, \he3/H, and \li7/H
are perhaps most clear in the bottom panel of the figure. 
Qualitatively,
we see that D decreases while \he3 and \li7 are both increased.  
This is the expected result of the boost in \he3 production at
the expense of D, and then subsequent $\he3(\alpha,\gamma)\be7$ production. 

Quantitatively, we find that at $\eta_{10} = 6.129$,
the shifts to (D/H,\he3/H,\li7/H) are $(-6.1\%, +7.6\%, +11.8\%)$, respectively.  These are larger 
than the shifts we have seen due to the \be7 rates.  
Of course the change in \he3 is inconsequential given the current
lack of observational data.  The increase in \li7/H makes the
lithium problem somewhat {\em worse}, and the effect of this
reaction is a large part of why our lithium prediction differs
from that of Pitrou et al.~\cite{coc18}.  
But the systematic shift in D/H is significant compared to the 
D/H prediction uncertainty we have quoted, and it is even more significant compared to the $\sim 1$\% errors in the observations.

Indeed, such a shift in D/H will spoil the excellent agreement 
with observations that we have seen above.  Such a shift away from the present data would suggest that the 
quoted experimental errors are underestimated, so the statistical
significance of such a shift would need to reflect a
larger error budget than we (and others) have contemplated.
Alternatively, such a shift would imply that the observational
uncertainties have been underestimated. In the extreme, the shift could imply physics beyond the standard model which
alters SBBN. We emphasize, however, that current nuclear experimental data and observations do not imply any departure from SBBN.

Clearly this is an important issue, as has been emphasized
by several groups \cite{cookeN,bbnt,Nollett2011,coc15,DiV}.
It is thus of the highest importance that 
new measurements be done for this reaction
at BBN energies.  We eagerly await these results.

\section*{Acknowledgments}
We are grateful to our longtime collaborator
Richard Cyburt for discussions
on this and related work.
We are also pleased to thank Lloyd Knox and Joaquin Vieira
for useful conversations regarding the CMB and 
future Stage 4 capabilities.
And we remember our friend an colleague Gary Steigman
whose thinking informs much of our own.
We acknowledge the use of {\tt COSMOMC} \cite{cosmomc}
to prepare Figs.~\ref{fig:triangle-TTTEEEnolensing} and
\ref{fig:triangle-TTTEEEcompare}.
The work of B.D.F. and T.-H. Y. was partially supported by the National Science Foundation Grant No. PHY-1214082.
The work of K.A.O.~was supported in part by DOE grant DE-SC0011842  at the University of
Minnesota.
K.A.O.
 acknowledges support by the Director, Office of Science, Office of High Energy Physics of the U.S. Department of Energy under the Contract No. DE-AC02-05CH11231.
K.A.O. would also like to thank the Department of Physics and the 
high energy theory group
at the University of California, Berkeley as well as the theory group at LBNL
for their hospitality and financial support while
finishing this work. 
B.D.F.~benefited from discussions with the participants of the 2019 Lithium in the Universe conference, and at the 2019 Frontiers of Nuclear Astrophysics conference, supported by the National Science Foundation under Grant No. PHY-1430152 (JINA Center for the Evolution of the Elements).

\end{document}